\documentclass[reqno,12pt]{article}
\usepackage[usenames,dvipsnames]{xcolor}
\usepackage{ae}
\usepackage[T1]{fontenc}
\usepackage[ansinew]{inputenc}
\usepackage{mathrsfs}
\usepackage{amsmath}
\usepackage{amssymb}
\usepackage{graphicx}
\usepackage{color}
\definecolor{darkblue}{cmyk}{0.9,0.9,0,0}
\definecolor{darkgreen}{rgb}{0,0.55,0}
\usepackage[colorlinks=true,linkcolor=darkblue,citecolor=darkblue,urlcolor=darkblue]{hyperref}
\usepackage{epsfig}
\usepackage{turnstile}
\usepackage{bbm}

\newcommand{\comment}[1]{}

\newcommand{\beq}{\begin{equation}}
\newcommand{\eeq}{\end{equation}}
\newcommand{\beqq}{\begin{equation*}}
\newcommand{\eeqq}{\end{equation*}}
\newcommand\beqa{\begin{eqnarray}}
\newcommand\eeqa{\end{eqnarray}}
\newcommand\beqaa{\begin{eqnarray*}}
\newcommand\eeqaa{\end{eqnarray*}}
\newcommand\bea{\begin{array}}
\newcommand\eea{\end{array}}

\def\XXint#1#2#3{{\setbox0=\hbox{$#1{#2#3}{\int}$}
\vcenter{\hbox{$#2#3$}}\kern-.5\wd0}}

\newcommand{\nn}{\nonumber}

\newcommand{\neqa}{\nonumber\end{eqnarray}}
\newcommand{\la}[1]{\label{#1}}

\newcommand{\eq}[1]{(\ref{#1})}

\newcommand{\Tr}{{\rm Tr}}

\renewcommand{\d}{\partial}

\newcommand{\<}{{\langle}}
\renewcommand{\>}{{\rangle}}

\newcommand{\cckappa}[4]{\Big(\!\!\!\Big({\color{OliveGreen}#3}\Big]\!\!
\Big[
{\color{OliveGreen}#4}\Big)\!\!\!\Big)_{#1}^{#2}}
\newcommand{\bv}{\overline{{\rm v}}}

\newcommand{\ckappa}[3]{\Big(\!\!\!\Big({\color{MidnightBlue}#3}\Big)\!\!\!\Big)_{#1}^{#2}}

\newcommand{\re}{\relax{\rm I\kern-.18em R}}

\renewcommand{\sp}{p\hspace{-.40em}/}

\def\su2{{SU(2)}}

\def\[{\left[}
\def\]{\right]}

\def\({\left(}
\def\){\right)}
\def\[{\left[}
\def\]{\right]}

\def\<{\langle}
\def\>{\rangle}

\def\i2{\frac{i}{2}}

\def\O{{\mathcal O}}

\def\spi{\relax{\rm \pi\kern-0.5em /}}
\def\sA{\relax{\rm A\kern-0.5em /}}
\def\sp{\relax{\rm p\kern-0.5em /}}
\def\sd{\relax{\rm \d\kern-0.5em /}}
\def\sk{\relax{\rm k\kern-0.5em /}}
\def\sn{\relax{\rm n\kern-0.5em /}}
\def\sl{\relax{\rm l\kern-0.5em /}}
\def\sP{\relax{\rm P\kern-0.7em /}}
\def\sBethe{\relax{\rm \Bethe\kern-0.5em /}}

\def\bu{{\bf u}}
\def\bv{{\bf v}}

\usepackage{varioref}
\usepackage{makeidx}
\makeindex

\usepackage[english]{babel}
\begin{document}

\thispagestyle{empty}

\renewcommand{\thefootnote}{\fnsymbol{footnote}}
\setcounter{page}{1}
\setcounter{footnote}{0}
\setcounter{figure}{0}
\begin{center}
$$$$
{\Large\textbf{\mathversion{bold}
Tailoring Three-Point Functions and Integrability IV. \\
$\Theta$--morphism
}\par}

\vspace{1.0cm}

\textrm{Nikolay Gromov$^{\,\theta_1}$ and Pedro Vieira$^{\,\theta_L}$}
\\ \vspace{1.2cm}
\footnotesize{
\textit{$^{\theta_1}$ King's College London, Department of Mathematics WC2R 2LS, UK \& \\
St.Petersburg INP, St.Petersburg, Russia} \\
\texttt{nikgromov@gmail.com} \\
\vspace{3mm}
\textit{$^{\theta_L}$ Perimeter Institute for Theoretical Physics\\ Waterloo,
Ontario N2L 2Y5, Canada}  \\
\texttt{pedrogvieira@gmail.com}
\vspace{4mm}
}

\par\vspace{1.5cm}

\textbf{Abstract}\vspace{2mm}
\end{center}
We compute structure constants in $\mathcal{N}=4$ SYM at one loop using Integrability. This requires having full control over the two loop eigenvectors of the dilatation
operator for operators of arbitrary size. To achieve this, we develop an algebraic description
called the $\Theta$--morphism.
In this approach we introduce impurities at each spin chain site,
act with particular differential operators on the standard algebraic Bethe ansatz vectors and
generate in this way higher loop eigenvectors.
The final results for the structure constants take a surprisingly simple form. For some quantities we conjecture all loop generalizations. These are based on the tree level and one loop patterns together and also on some higher loop experiments involving simple operators.
\noindent

\setcounter{page}{1}
\renewcommand{\thefootnote}{\arabic{footnote}}
\setcounter{footnote}{0}

 \def\nref#1{{(\ref{#1})}}

\newpage
\tableofcontents

\newpage
\section{Introduction}
One of the main reasons for trying to solve a gauge theory such as $\mathcal{N}=4$ SYM is to search for hidden structures pointing towards alternative descriptions of Quantum Field Theory. { As a byproduct, we also hope to develop new techniques that will allow us to do computations which are currently beyond reach and which will improve our understanding of String Theory and Quantum Gravity.}

To illustrate how such structures are often hidden we will consider the problem of computing correlation functions of local gauge invariant operators in the planar limit of $\mathcal{N}=4$ SYM.
For simplicity, let us consider operators made out of two complex scalars only, say $Z$ and $X$. As usual we identify them as spins for notational convenience,
\beq
\Tr (ZX\dots) \leftrightarrow \left|\uparrow\downarrow\dots\right\> \,.
\eeq
We use $N$ for the number of $X$ fields and $L$ for the total number of fields. Then, the gauge invariant operators with definite total dimension $\Delta=L+\gamma$ are the eigenvectors of the
dilatation operator which can be written as a spin chain Hamiltonian as \cite{MZ,BKS}
\beq
H=\sum_{n=1}^L \[ (2g^2-8g^4) \, \mathbb{H}_{n,n+1} +2 g^4\,  \mathbb{H}_{n,n+2}  + \mathcal{O}(g^6)\] \la{dilatation0}\;.
\eeq
Here $\mathbb{H}_{nm}=\mathbb{I}_{nm}-\mathbb{P}_{nm}$, with $\mathbb{I}$, $\mathbb{P}$ the identity and permutation operator, respectively. The energies of
this Hamiltonian are precisely the anomalous part of the total dimension, $\gamma$.
We can already identify remarkable simplifications which come about when considering physical quantities but which are very much obscured in the intermediate steps:
take for example the dilatation operator (\ref{dilatation0}). If we were to write its next few loop orders it would require pages and pages with no obvious structure present. Nevertheless, its eigenvalues, which are the physical quantities $\gamma$  can be obtained very neatly as we now briefly review. We can think of the states diagonalizing (\ref{dilatation0}) as $N$ excitations $\downarrow$ moving with momenta ${\bf p}=\{p_1,\dots,p_N\}$ on top of a ferromagnetic sea of $\uparrow$ fields. The scattering of two magnons is governed by the magnon S-matrix $S(p_i,p_j)$.\footnote{{ The scattering of $N$ particles is also governed by the two magnon S-matrix since
the many body S-matrix factorizes into a product of two body scattering processes in an Integrable theory.}} The quantization of these momenta is given by Bethe equations\footnote{See section \ref{contactSec} for a very brief review and \cite{paper1} for a more extended one.}
\beq
L p_k + \sum_{l \neq k} \frac{1}{i} \log S(p_l,p_k) = 2\pi n_k \,,\qquad n_k \in \mathbb{Z} \,. \la{BetheEq0}
\eeq
The S-matrix is known to all loop orders \cite{Beisert,BES} and is quite simple. Once the momenta are fixed the energy of the state is obtained by summing the energies $\epsilon(p_j)$ of each individual excitation,
\beq
\gamma({\bf p})= \sum_{j=1}^N \epsilon(p_j)\,, \qquad \epsilon(p)=8g^2 \sin^2(p/2)-32 g^4\sin^4(p/2)+\mathcal{O}(g^6) \la{gamma0} \,.
\eeq
This scattering picture needs to be corrected if the chains are too small \cite{Matthias,TBA}. That is, the Bethe ansatz (\ref{BetheEq0}) is asymptotic.  Equations (\ref{gamma0}) and (\ref{BetheEq0}) are a striking example of how beauty emerges at the end of the day for physical quantities.

Once the dimensions of operators are known we know their two point functions. The next natural observables are the structure constants which govern the three point functions. For three point functions we also need the eigenvectors of the dilation operator.
 After all we need to Wick contract them (at leading order) and then decorate this contraction by loops. However, the eigenvectors can be
 horrendous at first sight.
 More specifically, at two loops (and higher) there will be some local corrections to the wave functions, called \textit{contact terms}, when two or more magnons are close together. When four excitations are close together the corresponding two loop contact term correction is as messy as to occupy one full page!, see appendix \ref{C4}.

For the spectrum problem we could avoid dealing with these contact terms by considering asymptotically large chains where there is always enough
room for the excitations to move freely. Then, to quantize the momenta of the excitations all we care about is the S-matrix and to compute their energy we can consider a region where they are well separated. However, when we move to three point functions
the contact terms come into play. When we contract the three operators we need to know the exact wave function. All regions will be probed and all of them will contribute. On the other hand,
the structure constants $C_{123}$ are physical quantities and as such, there ought to be some nice description where all the complications due to the contact terms and the loop corrections conspire together to give a neat final result.

Confirming this expectation is the main goal of the current paper. To arrive at this result we propose an algebraic description of the eigenvectors of the dilation operator (\ref{dilatation0}) which automatically incorporates all possible contact terms. This approach is based on the so called $\Theta$--morphism which gives the name to the paper.

Let us end this introduction by mentioning that there are numerous other exciting hints at hidden structures, other than the $\Theta$--morphism \cite{BDS,boost,Hubbard,sigma,didina}. They strongly inspired this work.
It would be wonderful to explore in detail the different connections and differences between all these approaches.

This paper is organized as follows. In section \ref{setup} we translate the problem of computing structure constants into the spin chain language. The rest of the paper is divided into two main parts. In section \ref{coorSec} we follow the coordinate Bethe ansatz approach combined with some inspired guesses based on our experience with the spectrum problem. This section will contain several conjectures for the one loop correction to the structure constants and also for some all loop generalizations. In sections \ref{algSec} and \ref{secMor1} we follow an algebraic Bethe ansatz approach and we develop the $\Theta$--morphism map. This will allow us to derive the most general $SU(2)$ structure constants and prove all the one loop conjectures of section \ref{coorSec}, based on the coordinate Bethe ansatz approach.  We conclude in section \ref{Conc}. Several appendices contain complementary material to the main text.

\section{Structure Constants and Quantum Spin Chains} \la{setup}
In this paper we will consider structure constants in the $SU(2)$ setup introduced in \cite{paper1}.
Let us recall how these objects can be efficiently computed using the spin chain picture. At tree level all we need to do is to Wick contract the three operators. These operators are the eigenvectors of the dilatation operator (\ref{dilatation0}).
At one loop level (a) we need to Wick contract the operators which are the eigenvectors of the two loop dilatation operator  and (b) we need to include Hamiltonian insertions coming from Feynman diagrams
\cite{C123p1,C123p2,C123p3}
(see figure \ref{figure}). In total
\begin{figure}[t]
\begin{center}
\includegraphics[width=140mm]{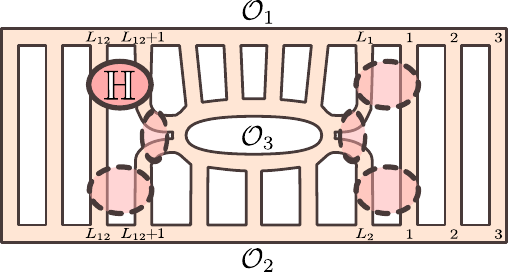}
\end{center}
\caption{Internal loop diagrams results in the insertions of the Hamiltonian densities at $6$ junction points \cite{C123p1,C123p2,C123p3}. This leads to (\ref{C123}) and (\ref{tocompute}) in the spin chain language.} \la{figure}
\end{figure}
\beqa
C_{123}&=&\verb"norms" \times \verb"simple" \times \verb"involved" +\mathcal{O}(g^4) \la{C123}
\eeqa
where
\beqa
\verb"norms" &=&{\sqrt{L_1L_2L_3}}/{\sqrt{\<{\bf 1}|{\bf 1}\>\<{\bf 2}|{\bf 2}\>\<{\bf 3}|{\bf 3}\>}} \,, \nn \\
\verb"simple" &=&\<\underbrace{\downarrow\dots\downarrow}_{N_3} \underbrace{\uparrow\dots\uparrow}_{L_3-N_3}|1-g^2 \mathbb{H}_{N_{3},N_{3}+1}-g^2 \mathbb{H}_{L_3,1} |{\bf 3}\> \la{tocompute}\,,\\
\verb"involved" &=&\<{\bf 1}|1-g^2 \mathbb{H}_{L_{12},L_{12}+1}-g^2 \mathbb{H}_{L_1,1} \nn
|
{{i_1}\dots{i_{L_{12}}}}
\underbrace{\downarrow\dots\downarrow}_{N_3}\>\\
\nn&&\<
{{i_1}\dots {i_{L_{12}}}}
\underbrace{\uparrow\dots\uparrow}_{L_3-N_3}|
1-g^2 \mathbb{H}_{L_{12},L_{12}+1}-g^2 \mathbb{H}_{L_2,1}|{\bf 2}\>\,.
\eeqa
Where $i_a$ can be $\uparrow$ or $\downarrow$ and one sums over all possible $2^{L_{12}}$ intermediate states.
Here $|\bf a\>$ stands for the corrected two loop Bethe eigenstate corresponding to the operator $\mathcal{O}_a$. The length and number of excitations of this operator are denoted by $L_a$ and $N_a$ respectively.
Of course, for these expressions to make sense and to be non zero we need
\beq
N_1 = N_2 + N_3 \qquad \text{and} \qquad L_{12} = L_1-N_3=L_2-L_3+N_3 \la{constraint}
\eeq
where $L_{12}$ is the number of contractions between operators $\O_1$ and $\O_2$. For any pair of operators $\O_i$ and $\O_j$ we have $2L_{ij}=L_i+L_j-L_k$ where $k\neq i,j$.

The computation of structure constants in $\mathcal{N}=4$ SYM in the $SU(2)$ sector up to one loop order is thus mapped into a very precise spin chain problem:
we simply need to be able to compute all spin chain scalar products and expectation values appearing in (\ref{tocompute}). At order $g^0$ this was solved in \cite{paper1,omar}. One of the main goals of the current paper is to compute the first quantum correction, of order $g^2$.
This would be the first step toward an all loop asymptotic expression for the structure constants.
Obviously we want to find reasonable conjectures for all loop results based on the tree level and one loop results
based on our experience with the spectrum problem.
As an example, we will now consider the first contribution in (\ref{C123}), the $\verb"norms"$. For that,  we first need to work out the precise form of the two loop eigenvectors $|{\bf a}\>$ which will allow us to ask nonambiguous questions about their norms.

\section{Two Loop States and Contact Terms} \la{contactSec}
The dilatation operator in $\mathcal{N}=4$ SYM can be thought of as a long-range spin chain Hamiltonian whose interaction range coincides with the order of perturbation theory. At two loops we have (\ref{dilatation0}).
The single magnon state is given by
\beq
|p\> =\sum_{n=1}^L e^{ipn} |n\> \,, \qquad  |n_1,n_2,\dots \>= |\overbrace{\underbrace{\uparrow \dots \uparrow}_{n_1-1}\! \downarrow \uparrow \dots\uparrow}^{n_2-1}\! \downarrow \uparrow \dots \> = \sigma_-^{(n_1)}\sigma_-^{(n_2)}\dots \left|\uparrow\dots \uparrow\right\>\;.
\eeq
This follows from translation invariance and is true to all orders in perturbation theory. The two magnon state is more interesting. At two loops we have
\beq
|p_1,p_2\> =\sum_{n_1 < n_2} \(e^{ip_1n_1+ip_2n_2}+e^{ip_2n_1+ip_1n_2} S(p_2,p_1) \)\(1+ \delta_{n_2,n_1+1} \,g^2\, \mathbb{C}_{\bullet\bullet}^{(2)}(p_2,p_1) \) |n_1,n_2\> \la{2magnon}\;.
\eeq
The physical picture is clear: we have an incoming wave ($e^{ip_1n_1+ip_2n_2}$) describing two particles with momentum $p_1$ and $p_2$. They can then collide. In $1+1$ dimensions when two particles collide their momentum can at most be exchanged. This is a consequence of conservation of both energy and momentum. The outgoing plane wave ($e^{ip_2n_1+ip_1n_2}$) describes the particles with their momenta exchanged. The relative coefficient between the incoming and outgoing plane waves is what we define as the S-matrix, which turns out to be given by \cite{BDS,Matthias,BES}
\beq
S(p,k)= \frac{u(k)-u(p)-i}{u(p)-u(k)+i} \,, \qquad u(p)=\frac{1}{2}\cot\(\frac{p}{2}\) \(1+ 8g^2 \sin^2 \frac{p}{2} + \dots \) \la{Smatrix}\;.
\eeq
This is all we have as long as the particles are well separated. When they are within the interaction range of the Hamiltonian the two magnons will be in a complicated interacting state. This is taken into account by introducing the so called \textit{fudge factors} \cite{Matthias} which we also denote as \textit{contact terms}. The purpose of this section is to study these terms. At two loops, and for two magnons, we only need
\beq\mathbb{C}_{\bullet\bullet}^{(2)}(p_2,p_1) = -4 \sin(p_1/2) \sin(p_2/2)\sec((p_1+p_2)/2)  \la{C2}\,.\eeq
The superscript indicates that this is a two loop effect while the subscript indicates that the two magnons need to be close to each other.

The higher loop picture should be clear; new fudge factors have to be added with longer and longer range. For example, at three loops, the two magnon the state (\ref{2magnon}) is modified by the simple replacement of $\(1+ \delta_{n_1+1,n_2}\,g^2\, \mathbb{C}_{\bullet\bullet}^{(2)}(p_2,p_1) \) $ by
\beq
 \Big(1+ \delta_{n_2,n_1+1} \,\[g^2\, \mathbb{C}_{\bullet\bullet}^{(2)}(p_2,p_1) +g^4\, \mathbb{C}_{\bullet\bullet}^{(3)}(p_2,p_1)\]+ \delta_{n_2,n_1+2} \,g^4\, \mathbb{C}_{\bullet\circ\bullet}^{(3)}(p_2,p_1) \Big) \,.
\eeq
Also, the S-matrix needs to be corrected further and at four loop order a new contribution to the S-matrix -- the so called BES dressing phase \cite{BES} -- shows up. For now let us continue our two loop discussion but move to a higher number of excitations.

Interesting new structures arise when we move to three (and more) magnons. For three magnons we have
\beq|p_1,p_2,p_3\> =\sum_{n_1 < n_2<n_3} \!\!\!\![\phi_{123}+\phi_{213} S_{21}+\phi_{132} S_{32} +\phi_{312} S_{31}  S_{32} +\phi_{231}S_{31} S_{21}+\phi_{321} S_{32}  S_{31}  S_{21}] |n_1,n_2,n_3\>\nn\eeq where $S_{ab}=S(p_a,p_b)$ and
\beq \phi_{abc}= \exp( ip_a n_1+i p_b n_2+ i p_c n_3) (1+g^2 \delta_{abc})\,.\eeq
Finally -- and most relevant for the current discussion -- $\delta_{abc}$ are the contact terms. We have
\beq
\delta_{abc} =  \left\{ \begin{array}{ll} 0 & \text{for  } n_2-n_1>1 \text{ and } n_3-n_2>1  \\
 \mathbb{C}_{\bullet\bullet}^{(2)}(p_a,p_b) &   \text{for  } n_2-n_1=1 \text{ and } n_3-n_2>1 \\
 \mathbb{C}_{\bullet\bullet}^{(2)}(p_b,p_c) &   \text{for  } n_2-n_1>1 \text{ and } n_3-n_2=1 \\
 \mathbb{C}_{\bullet\bullet\bullet}^{(2)}(p_a,p_b,p_c) & \text{for  } n_2-n_1=1 \text{ and } n_3-n_2=1 \\
 \end{array} \right.
\eeq
We see that on top of the two magnon contact term we now need a new contact term for the case when the three magnons are all close to each other.
States with more magnons follow the same pattern. For $N$ magnons we will have $N!$  terms in the wave function and $N-1$ different contact terms: $\mathbb{C}^{(2)}_{\bullet\bullet}(p_1,p_2), \mathbb{C}^{(2)}_{\bullet\bullet\bullet}(p_1,p_2,p_3), \dots, \mathbb{C}^{(2)}_{\bullet\dots\bullet}(p_1,\dots,p_N)$.
By explicit diagonalization of the two loop dilatation operator we can compute the first few contact terms. We find
\beqa
\mathbb{C}_{\bullet\bullet}^{(2)}(p_1,p_2)\!\!\!&=&\! \!\! \text{see equation (\ref{C2})} \nn \\ 
\mathbb{C}_{\bullet\bullet\bullet}^{(2)}(p_1,p_2,p_3)\!\!\!&=&\! \!\!\frac{2 (\cos (p_1+p_2)+\cos (p_1+p_3)+\cos (p_2+p_3)-2 \cos (p_1+p_2+p_3)-1)}{1-\cos (p_1)-\cos (p_2)-\cos
   (p_3)-\cos (p_1+p_2+p_3)} \nn \\
\mathbb{C}_{\bullet\bullet\bullet\bullet}^{(2)}(p_1,p_2,p_3,p_4)\!\!\!&=&\!\!\! \text{see appendix \ref{C4}} \la{contactTerms}
\eeqa
The complexity seems to grow in a scary and uncontrollable fashion. As we will derive later through indirect means, all the contact terms can be written very explicitly as ratios of complicated sums, see appendix \ref{codeContact}. In practice for $N>4$ it takes forever to compute these sums analytically even using a computer (numerically one can easily go up to much larger $N$'s).
For now, let us continue our experimental investigation and turn our attention to more physical quantities such as the structure constants,
in the hope that the huge mess in the contact terms will somehow miraculously drop out. As argued in the introduction this should be the case and indeed it will be.

Finally, let us end by recalling that the momenta of these Bethe eigenstates is quantized through the Bethe equations (\ref{BetheEq0}) which follows from the periodicity of the many particle wave function. The energy of the states is given by (\ref{gamma0}).
\section{Coordinate Bethe Ansatz. Educated Guesses. } \la{coorSec}

To compute one loop structure constants we need control over the spin chain scalar product and expectation values such as the ones in (\ref{tocompute}). We know them at tree level, that is for the one loop eigenvectors of the Dilatation operator. The purpose of this section is to guess their higher loop generalization by doing experiments. In particular, these experiments will allow us to
find an analytic expression for the simplest nontrivial structure constant (corresponding to the case when $\O_1$ is BPS).

We denote the momenta of the Bethe roots of $\O_1$, $\O_2$ and $\O_3$ as ${\bf k}$, ${\bf p }$ and ${\bf q}$ respectively.  Sometimes we shall use $|{\bf 2}\>$ and sometimes we will instead write $|{\bf p}\>$ etc.

\subsection{All Loop Norm Conjecture} \la{normsec}
The simplest scalar products appearing in our list (\ref{tocompute}) are the norms of the Bethe eigenstates.\footnote{Usually a norm is  not a physical object; it depends on the normalization of the state.
In the case of the Bethe sates, however, the normalization is fixed uniquely by the clear physical picture of scattering of the magnons.
Namely, we normalize the state by fixing to $1$ the coefficient of the incident wave, see e.g. (\ref{2magnon}).
After this the norm is well defined and is entitled with some physical meaning which is to be uncovered.}
At one loop the norm of a Bethe eigenstate is given by Gaudin's determinant \cite{Gaudin,paper1}
\beq
\<{\bf p}| {\bf p}\>_\text{1 loop}=\[\prod_{i<j} \frac{S^{(0)}(p_i,p_j)}{S^{(0)}(p_i^*,p_j^*)}\]^{1/2} \det_{1\le j,k \le N} \[\frac{\partial}{\partial p_j} \( L p_k + \sum_{l \neq k} \frac{1}{i} \log S^{(0)}(p_l,p_k) \)\]  + \mathcal{O}(\lambda) \la{Norm0}
\eeq
where ${\bf p }=\{p_j\}$ are the momenta of the $N$ magnons and $L$ is the total length of the operator. At one loop the S-matrix is given by the leading order in (\ref{Smatrix}) which we denoted by $S^{(0)}$. Finally $p_i^*$ is the
complex conjugate of $p_i$; the first factor in (\ref{Norm0}) is absent for real momenta.  Note that the expression inside the determinant is nothing but the left-hand side of Bethe equations (\ref{BetheEq0}) at leading one loop order.
A natural guess for the higher loop generalization is that we simply have to correct the S-matrix,
\beq
\<{\bf p}| {\bf p}\>_\text{all loops}\stackrel{?}{=} \[\prod_{i<j} \frac{S(p_i,p_j)}{S(p_i^*,p_j^*)}\]^{1/2}  \det_{1\le j,k \le N} \[\frac{\partial}{\partial p_j} \( L p_k + \sum_{l \neq k} \frac{1}{i} \log S(p_l,p_k) \)\]   \la{Norm}\;.
\eeq
At  first  this guess may look oversimplified.
After all, when going to higher loops we also needed to include the contact terms (\ref{contactTerms}) which were quite complicated
and one could suspect that the above result is correct modulo these contact terms.
A simple way to settle this question is to perform some {experiments}.\footnote{Latter we will derive (\ref{Norm}) for two loops from an algebraic Bethe ansatz approach where no guesswork will be involved.}
We computed the norm for several different eigenstates with different number of magnons by brute force up to $4$ loops.
Remarkably, we find that (\ref{Norm}) works perfectly when the contact terms are included!
On the contrary, it is \textit{because of the very complicated contact terms} that the final result turns out to be so remarkably simple!
This clearly indicates that some deeper structure must be hidden; unveiling it (at least at two loops) will be the purpose of later sections.
Remarkably, at 4 loops the dressing phase is nontrivial and affects the eigenstate.
Nevertheless we find that (\ref{Norm}) works perfectly! 
This leads to the
conjecture:

\textit{For large enough operators, in the usual asymptotic sense, (\ref{Norm}) holds to all loops. }

\subsection{Double Vacuum Decay Amplitudes and Structure Constants}
We now have control over the \verb"norms" contribution in (\ref{C123}), see previous section. We will now move to the remaining contributions: \verb"simple" and \verb"involved". In this section we will
\begin{itemize}
\item Compute \verb"simple"
\item Compute \verb"involved" \textit{for} the simpler cases where $\O_1$ is a BPS operator.
\end{itemize}
It turns out that for $\O_1$ BPS, the contribution \verb"involved" is as simple as the \verb"simple" term. More precisely, these quantities turn out to be related to an important scalar product which we call the Double Vacuum Decay (DVD) Amplitude, see below. Guessing the higher loop expression for this scalar product is the main goal of the current section.

\subsubsection{The Double Vacuum Decay (DVD) Amplitude} \la{simplificationSec}
We will now introduce an important building block of the structure constants which we call Double Vacuum Decay (DVD) amplitude.

To define it let us describe zero energy states of the dilatation operator. We will work with a particular basis of zero energy states which are the so called vacuum descendents of length $L$ and $N$ spin flips.
We denote them as
\beq
\left|\,\ssststile{\,\,\,\,L\,\,\,\,}{\,\,N\,\,}\,\right\> \equiv
|\! \underbrace{\uparrow\dots\uparrow}_{L-N} \underbrace{\downarrow\dots\downarrow}_{N}\,\>+\text{all $\binom{L}{N}$ permutations}\;.
\eeq
We can glue two such vacuum states one after the other making what we call a Double Vacuum $\left|\,\ssststile{\,\,\,\,L\,\,\,\,}{\,\,N\,\,}\,\,\ssststile{\,\,\,\,L'\,\,\,\,}{\,\,N'\,\,}\,\right\>\equiv \left|\,\ssststile{\,\,\,\,L\,\,\,\,}{\,\,N\,\,}\,\right\> \otimes \left|\,\ssststile{\,\,\,\,L'\,\,\,\,}{\,\,N'\,\,}\,\right\>$.
The DVD amplitude is defined as a scalar product between such double vacuum and a Bethe state
\beq
\mathcal{A}_{L}({\bf q})\equiv (-1)^N\left\<\,\ssststile{\,\,\,\,L\,\,\,\,}{\,\,N'-N\,\,}\,\ssststile{\,\,\,\,L'-L\,\,\,\,}{\,\,N\,\,}\Big| \,{\bf q}\right\> \la{Adef}
\eeq
Here $L'$ is the length of the Bethe eigenstate $|{\bf q}\>$ and $N'$ is the number of magnons in it.
If all momenta $\bf q$ are nonzero the Bethe eigenstate is a highest weight state. In this case the DVD amplitude does not depend on $N$.\footnote{
The Bethe state is annihilated by $S^+$. At the same time the adjoint operator $S^-$
acting on the bra state would mix states with two various $N$ from which we conclude that they must be equal up to a sign.}

The DVD amplitude (\ref{Adef}) is an important building block for the structure constants as we will now illustrate.
For example,
the \verb"simple" contribution in (\ref{tocompute})
in the new notation becomes
\beq
\verb"simple"=\left\<\,\ssststile{\,\,\,\,N_3\,\,\,\,}{\,\,N_3\,\,}\,\ssststile{\,\,\,\,L_3-N_3\,\,\,\,}{\,\,0\,\,}\,\right|1-
g^2 \( \mathbb{H}_{N_{3},N_{3}+1}+ \mathbb{H}_{L_3,1}\) \Big|{\bf 3}\Big\>\la{tocompute2} \,.
\eeq
Note that the bra state is annihilated by  $\mathbb{H}_{i,i+1}$ unless $(i,i+1)=(N_3,N_3+1)$ or $(L_3,1)$. This means that we can replace $ \mathbb{H}_{L_{13},L_{13}+1}+ \mathbb{H}_{L_3,1}$ by $\sum_{i=1}^{L_3} \mathbb{H}_{i,i+1}$ in (\ref{tocompute2}). In this way we recognize the one loop dilation operator, see (\ref{dilatation0}), up to a factor of $1/2$.  Then we act with this operator to the right, on $|{\bf 3}\>$. At this loop order, we simply get (one half of) the energy of the Bethe eigenstate
\beq
\gamma({\bf q})= \sum_{j=1}^{N_3} 8 g^2 \sin^2\frac{q_j}{2} + \mathcal{O}(g^4) \la{Energy}
\eeq
where ${\bf q}=\{q_j\}$ are the momenta of the magnons of the operator $\mathcal{O}_3$.
We conclude that
\beq
\verb"simple"= \(1-\frac{1}{2} \gamma({\bf q})\) \mathcal{A}_{N_3}({\bf q})\;. \la{simplefinal}
\eeq

Physically, $\mathcal{A}_L$ describes the probability amplitude for all the $N$ magnons in $|{\bf q}\>$ to occupy homogeneously the first $L$ sites of the chain which has total length $L'>L$. Above $N=N_3$ and $L=N_3$ as well which means so that magnons are all squeezed together in the first $N_3$ sites.  The scalar product $\mathcal{A}_{L}({\bf q})$ was studied in detail in \cite{paper3} and will be considered again in section \ref{Asec}.

\subsubsection{Three Point Functions from DVD Amplitudes} \la{decaysec}
In the previous sections we computed the first contribution \verb"norms" in (\ref{C123}) and reduced the second contribution, \verb"simple", to the computation of the DVD amplitude $\mathcal{A}_{L}({\bf q})$ defined in (\ref{Adef}). We now turn our attention to the \verb"involved" contribution in (\ref{tocompute}).

In general, it is not possible to get rid of the Hamiltonian insertions in this quantity as easily as we did for the contribution \verb"simple". However, if $\mathcal{O}_1$ is a BPS state then we can still do it. In this case, we can follow the steps of the previous section to obtain
\beq
\verb"involved"=\(1-\frac{1}{2}\gamma({\bf p})\) \mathcal{A}_{L_1-N_3}({\bf p}) \la{hereitis}
\eeq
where ${\bf p}=\{p_1,\dots,p_{N_2}\}$ are the momenta of the magnons of the operator $\mathcal{O}_2$. For details see Appendix \ref{422ap}. Combining this result with (\ref{simplefinal}) we conclude that for $\mathcal{O}_1$ BPS we have
\beqa
\nn\text{$\O_1$ BPS:}\qquad C_{123}^{\circ\bullet\bullet}({\bf p},{\bf q})&=& {\sqrt{L_1L_2L_3}}{}
\(1-\frac{\gamma({\bf p})+\gamma({\bf q})}{2}\) \frac{ \mathcal{A}_{N_3}({\bf q})  \mathcal{A}_{L_1-N_3}({\bf p}) }{\sqrt{\binom{L_1}{N_1} \<{\bf 2}|{\bf 2}\>\<{\bf 3}|{\bf 3}\>}}  \la{C123oxx}\\
\text{ $\O_1$ and $\O_3$ BPS:}\qquad C_{123}^{\circ\bullet\circ}({\bf p},{\bf q})&=& {\sqrt{L_1L_2L_3}}
\(1-\frac{\gamma({\bf p})}{2}\) \frac{ \mathcal{A}_{L_1-N_3}({\bf p}) }{\sqrt{\binom{L_1}{N_1}\binom{L_3}{N_3}\<{\bf 2}|{\bf 2}\>}}  \la{C123oxo}\;.
\eeqa
We already found the norms in section (\ref{normsec}). We now turn to the DVD amplitude $\mathcal{A}$.
\subsubsection{Conjectures for the DVD Amplitude} \la{Asec}
In this section we find an explicit expression for the DVD Amplitude
defined in (\ref{Adef}) valid at tree level and at one loop.
It is the last missing piece to obtain the full one loop corrected structure constants of the type (\ref{C123oxx}).
 At tree level -- i.e. for one loop Bethe eigenstates --we have \cite{paper1}\footnote{As shown by Kostov, the sum (\ref{Asum}) can be recast as a determinant \cite{Ivan},
$$\sum\limits_{\alpha \cup \bar\alpha= \{p\}} (-1)^{|\alpha|} \prod\limits_{k \in \alpha, \bar k \in \bar \alpha} \! \! f(k,\bar k) \,\prod\limits_{\bar k \in \bar\alpha} e^{i L {\bar k}}=\det_{1\le j,k \le N} \[u(p_j)^{k-1}-e^{i p_j L} (u(p_j)-i)^{k-1}\]/\det_{1\le j,k \le N} \[u(p_j)^{k-1}\]\,.$$
 }
\beq
\mathcal{A}_{\text{tree level}}({\bf p})= \frac{\displaystyle\sum_{\alpha \cup \bar\alpha= \{p\}} (-1)^{|\alpha|} \! \! \prod_{k \in \alpha, \bar k \in \bar \alpha} \! \! f(k,\bar k) \,\prod_{\bar k \in \bar\alpha} e^{i L {\bar k}} }{\displaystyle\prod\limits_{j=1}^N \(e^{-ip_j}-1\) \prod\limits_{1\le i<j \le N} f(p_i,p_j)} \,. \la{Asum}
\eeq
The sum is over all possible partitions of the set of momenta and $|\bar\alpha|$ stands for the number of elements in the partition $\bar\alpha$.
For example, for two particles, we have
\beq
\mathcal{A}_{\text{tree level}}(p_1,p_2)= \frac{1 - f(p_1,p_2) e^{i L p_2}-f(p_2,p_1) e^{i L p_1}+e^{i L (p_1+p_2)} }{(e^{-ip_1}-1)(e^{-ip_2}-1)f(p_1,p_2)}  \la{A2}\;.
\eeq
Finally, $\displaystyle f(p,k)=1+i \[u(p)-u(k) \]^{-1}$
where $u(p)$ was defined in (\ref{Smatrix}). We will repeat the game of section \ref{normsec} and try to guess what the higher loop generalization of (\ref{Asum}) could be. The first natural guess is that this expectation value still takes the form (\ref{Asum}) where we simply need to correct $f(p,k)$. To check this proposal we computed (\ref{Adef}) for a two-loop corrected operator $\mathcal{O}_2$ with two magnons. For that we use the two loop wave functions, with contact terms, as described in the previous section, see appendix \ref{ApA1} for details. We realize that (\ref{Adef}) indeed still takes the form (\ref{A2}) with a corrected $f(p,k) \to \mathfrak f(p,k)$,
\beq
\mathfrak{f}(p,k)=\frac{u(p)-u(k)+i}{u(p)-u(k)} \(1+\frac{g^2}{(u(p)^2+\frac{1}{4})(u(k)^2+\frac{1}{4})} + \mathcal{O}(g^4)\) \,. \la{fcorr}
\eeq
We should emphasize that this is far from obvious. In fact, it is because of the particular form of the contact terms that the result can still be put in the form (\ref{A2}). By computing (\ref{Adef}) for a few magnons we verify that
\beq
\mathcal{A}_L({\bf p})= \frac{\displaystyle\sum_{\alpha \cup \bar\alpha= \{p\}} (-1)^{|\alpha|} \! \! \prod_{k \in \alpha, \bar k \in \bar \alpha} \! \! \mathfrak{f}(k,\bar k) \,\prod_{\bar k \in \bar\alpha} e^{i L {\bar k}} }{\displaystyle\prod\limits_{j=1}^N \(e^{-ip_j}-1\) \prod\limits_{1\le i<j \le N} \mathfrak{f}(p_i,p_j)}  \la{Afin}
\eeq
works perfectly at one loop! In fact, by appropriately correcting (\ref{fcorr}) we seem to be able to reproduce the two and three loop results \cite{GSV}. This leads to the next general conjecture:

\textit{For large enough operators, in the usual asymptotic sense, (\ref{Afin}) holds to all loops provided (\ref{fcorr}) is properly fixed. }\footnote{
At higher loops the Dilatation operator has scheme dependent terms which cannot be fixed by  symmetry and locality, see \cite{Beisert:2007hz}.
These terms have unfixed coefficients and do not affect the spectrum. They can be removed by a similarity transformation.
The DVD amplitude is not invariant under such transformations and can, in principle, contain some scheme dependence which of course should cancel out in physical quantities.
This subtlety will be considered elsewhere.}

For more details on how to fix $\mathfrak{f}(p,k)$ see appendix \ref{ApA2}. In particular, in this appendix, we present an alternative procedure which
constrains this function to all loop orders.

This concludes our study of the simplest correlation functions in (\ref{C123oxo}) where $\O_1$ is BPS. We conjectured the form of all the building blocks appearing in this expression and even proposed some all loop conjectures.

On the other hand we should conclude this section by mentioning that we failed in guessing most general
structure constants when both ${\cal O}_1$ and ${\cal O}_2$ are non-BPS operators.
This is partly because the tree level result simplifies enormously on-shell i.e. when
the set of roots ${\bf p}$ is constrained by the Bethe ansatz equations.
At the same time this complicates enormously the guesswork since this constraint
is rather complicated to deal with when doing experiments. In sum, the guessing strategy used so far seems insufficient for these more complicated cases.
We need some more scientific and systematic approach to find the one loop (and higher loops) deformation.
Developing such an approach is the main goal of the current paper and will be the focus of the remaining sections. The more impatient readers can check Appendix \ref{fullAfter} for the final  result for the most general one loop structure constants for three non-BPS operators.

\section{Algebraic Bethe Ansatz Approach and Impurities} \la{algSec}

In the previous section we followed the coordinate Bethe ansatz approach where the states are written as sums of plane waves with relative coefficients given by $S$-matrix elements. Beyond one loop we also have contact terms, see section \ref{contactSec}. In this section we will only consider the one loop states for which there is a well developed alternative approach known as the Algebraic Bethe ansatz. In this approach multi-particle states are simply given by
\beq
\includegraphics[trim=0cm 6cm 0cm 0cm, clip=true, scale=.6]{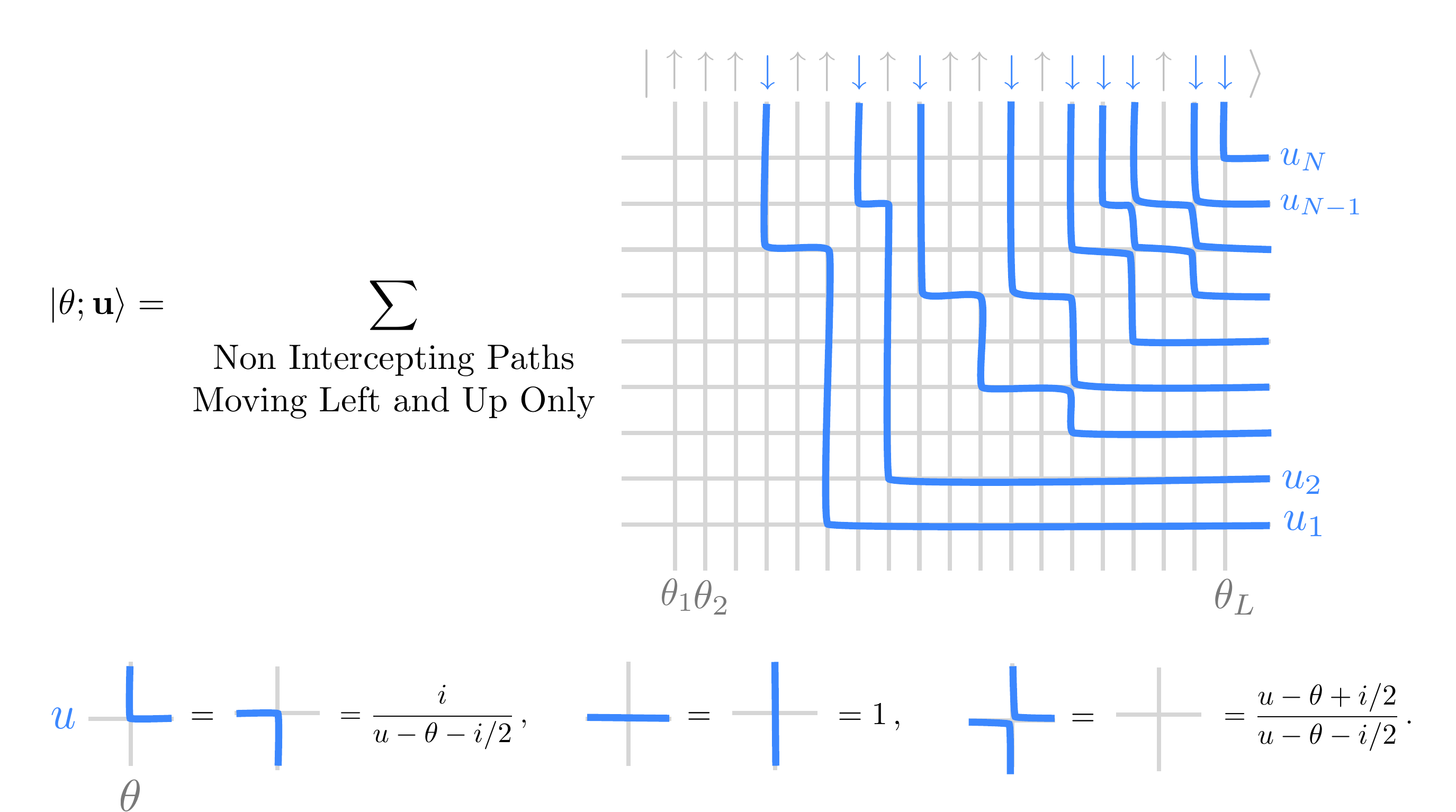} \la{stateAlg}
\eeq
In fact, the states represented here already contain a new twist: the impurities $\theta_j$ at each of the spin chain sites. To get the one loop Bethe eigenstates we set all impurities to zero, $\theta_j=0$ and consider $|0;{\bf u}\>$. However, as we will see in the next sections, these impurities will play a central role when going to higher loops.

The relation between the Bethe roots $u_j$ and the momenta $p_j$ is given by $u_j=u(p_j)$ where the change of variables was introduced in (\ref{Smatrix}).
This figure is a graphical implementation of the algebraic relation\footnote{More explicitly,
\beq
\hat B(u) = \left\<\uparrow\right| \bigotimes_{j=1}^L \(\mathbb{I}_{j0} +\frac{i}{u-\theta_j-i/2} \mathbb{P}_{j0} \)\left|\downarrow\right\>
\eeq
Here the subscript $j0$ indicates that the operator acts on the tensor product of physical Hilbert space $V_j = \mathbb{C}^2$ corresponding to spin chain site $j$ and an auxiliary space $V_0=\mathbb{C}^2$. The bra and ket in this formula live in this auxiliary space so that $\hat B(u)$ is an operator acting on the full spin chain Hilbert space $\mathcal{H}=V_1\otimes \dots \otimes V_L$. Graphically, the action of many such operators can be represented as in (\ref{stateAlg}). For algebraic Bethe ansatz reviews see for example \cite{paper1} and \cite{omar}.}
\beq
|\theta;{\bf u}\>=\hat B(u_1) \dots \hat B(u_N) \left|\uparrow\dots\uparrow\right\> \la{BBOmega}\;.
\eeq
It is quite remarkable that such creation operators $\hat B(u_j)$ exist at all. Even though  the different $\hat B$'s enter independently
they create fully interacting particles!
For example, for the two particle state, we obtain the following wave function (we have already set $\theta_j=0$ in this figure):
\beq
\includegraphics[trim=0cm 13cm 0cm 0cm, clip=true, scale=.6]{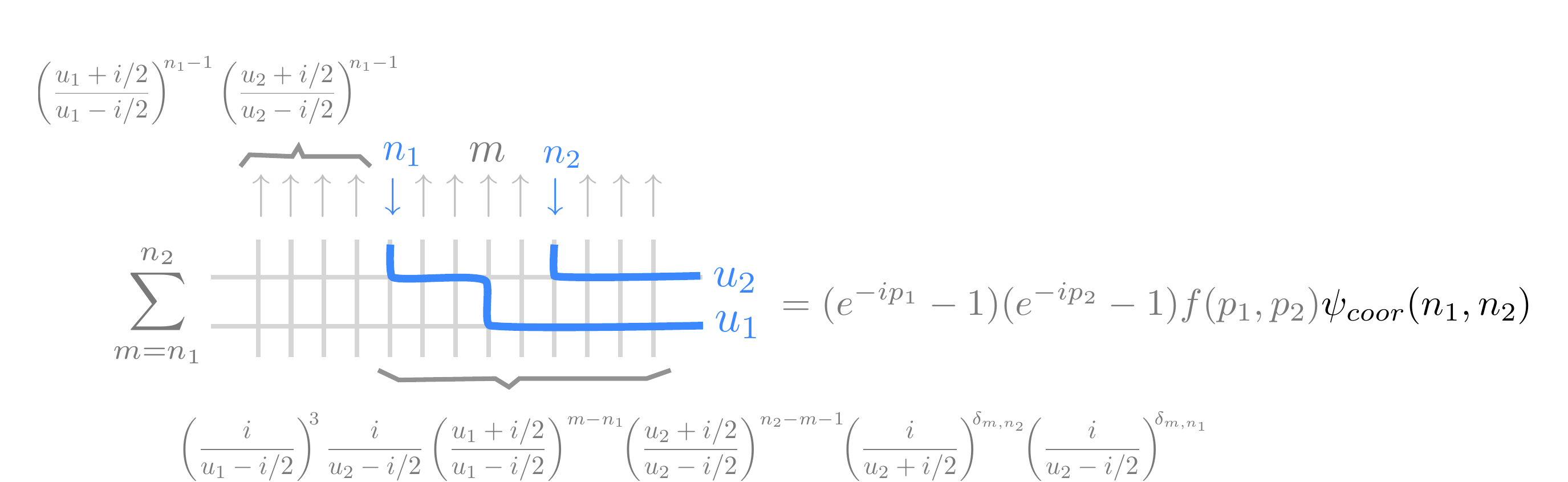} \la{2ptExample}
\eeq
which leads, up to a simple proportionality factor, to  the exactly the same wave function $\psi_{coor}(n_1,n_2)$ found through the Coordinate Bethe Ansatz approach,
\beq
\psi_{coor}(n_1,n_2)= e^{ip_1n_1+ip_2n_2}+S(p_2,p_1) e^{ip_2n_1+ip_1 n_2} \,,
\eeq
see (\ref{2magnon}) with $g^2=0$. Of course, this continues to hold for more magnons. For $N$ magnons, the relative normalization between the one loop coordinate states $|{\bf p}\>_\text{one loop}$ and algebraic states $|0;{\bf u}\>$ is given by
\beq
|0;{\bf u}\>=\mu\, |{\bf p}\>_\text{one loop} \,, \qquad \mu\equiv\prod_{j=1}^N (e^{-ip_j}-1) \prod_{j<k}^N f(p_j,p_k) \,. \la{lambda}
\eeq
where $f(p,k)=1+i/[u(p)-u(k)]$. 
Obviously, at the end of the day, all normalizations drop out for physical quantities.\footnote{As we emphasized
above the coordinate Bethe ansatz normalization is particular nice from a physical point of view
and we would like to be able to derive at least the first several loop orders for the norm. In (\ref{Norm}) we conjectured a value for the norm at all-loops in the coordinate Bethe ansatz normalization.}

\subsection{Impure States and Structure Constants} \la{impure}
To go to higher loops we need some sort of deformation of (\ref{BBOmega}) consistent with Integrability. The simplest one, which basically costs us nothing, is to introduce so called impurities $\theta_j$ at each lattice site $j$. In fact, in the previous section we already introduced these parameters, see (\ref{stateAlg}). At one loop we simply set those impurities to zero. The idea -- which will be explored in detail in the following sections -- is to use these impurities to mimic the loop corrected propagation and interaction of the magnon excitations. This deformation will amount to acting on the states (\ref{stateAlg}) with a differential operator made out of $\partial/\partial \theta_j$ and only then setting all impurities to zero at the end.

Concerning the structure constants, this procedure will allow us to promote tree level results with impurities to higher loop results without impurities. For future use let us quote here what the three point functions at tree level look like once we introduce the impurities, see figure \ref{figC123boxes}. This was worked out in \cite{omar}, see also \cite{short}.

Let us make a simplifying assumption that $\mathcal{O}_3$ is BPS. In any event, the contribution of $\mathcal{O}_3$ factorizes into a \verb"simple" contribution -- see section \ref{setup} -- so it is straightforward to generalize to the case with $\O_3$ non-BPS at any later stage if needed. 

We use $u_j=u(p_j)$ and $v_j=u(k_j)$ for the rapidities of operators  $\mathcal{O}_1$ and $\O_2$ respectively.

Finally we have to introduce some notation for the impurities. We denote the impurities of each operator $\O_a$ by $\theta_j^{(a)}$. Each impurity is associated to a line going from one operator to another so it is shared between two operators. We denote the set of impurities shared between $\O_a$ and $\O_b$ by $\theta^{(ab)}$ so that, for example, $\theta^{(2)}=\theta^{(12)}  \cup\theta^{(23)} $ and so on.
When omitting the superscript it is implicit that we refer to the first operator, $\theta_j\equiv \theta_j^{(1)}$. 

Then we have
\begin{figure}[t]
\begin{center}
\includegraphics[trim=9cm 1cm 0cm 0cm, clip=true, scale=.5]{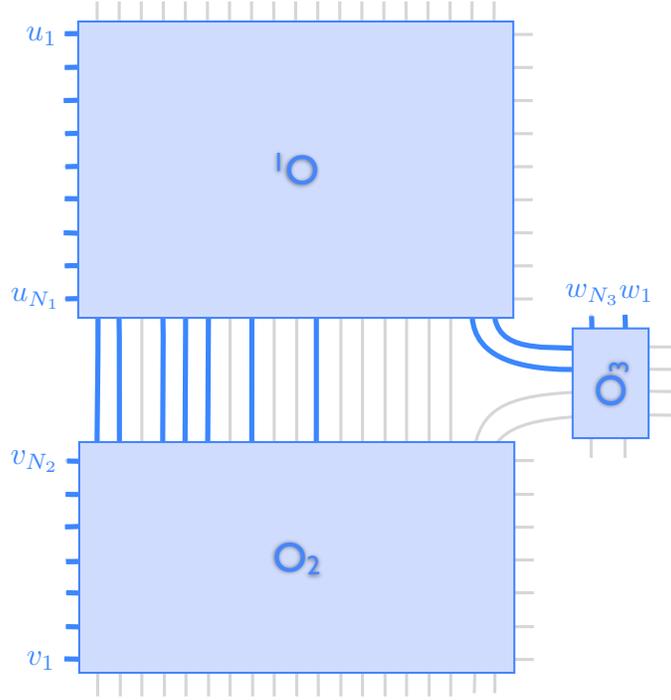}
\end{center}
\caption{Tree level impure structure constant. Each box represents a sum of paths as in (\ref{stateAlg}). We should also divide by the normalization of the three operators/spin chains. } \la{figC123boxes}
\end{figure}
\begin{equation}
|{C}^{\rm tree\;with\;imp.}_{123}| = \frac{\sqrt{L_1L_2L_3}}{\sqrt{\binom{L_3}{N_3}}}  \, \frac{|\<\theta^{(1)};{\bf u} | \hat{\mathcal{O}}_3| \theta^{(2)};{\bf v} \>|}{\sqrt{\left\<\theta^{(1)};{\bf u} |\theta^{(1)};{\bf u}\right\>\left\<\theta^{(2)};{\bf v} | \theta^{(2)};{\bf v} \right\>}}  \label{main} \,.
\end{equation}
where
\begin{equation}
\left\<\theta;{\bf u} |\theta;{\bf u}\right\>=\prod_{m\neq k} \frac{u_k-u_m+i}{u_k-u_m}   \det_{j,k\le N_1} \frac{\partial }{\partial u_j} \[ \frac{1}{i} \sum_{a=1}^{L_1} \log \frac{u_k-\theta_a+i/2}{u_k-\theta_a-i/2}
+\frac{1}{i}\sum_{m\neq k}^{N_1} \log\frac{u_k-u_m-i}{u_k-u_m+i}\]  \la{normB}
\end{equation}
with a similar expression for the other norm. Before presenting the numerator in (\ref{main}) let us discuss (\ref{normB}) briefly and make contact with the previous sections. The expression (\ref{normB}) is the norm of Bethe eigenstates with impurities. The Bethe equations with impurities take the form
\beq
 \sum_{a=1}^{L_1} \frac{1}{i} \log\(\frac{u_k-\theta_a+i/2}{u_k-\theta_a-i/2}\)+
\sum_{m\neq k}^{N_1} \frac{1}{i}\log\( \frac{u_k-u_m-i}{u_k-u_m+i}\) = 2\pi n_k \,, \qquad n_k \in \mathbb{Z} \la{BAEimpure}
\eeq
which reduces to (\ref{BetheEq0}) if we set all impurities to zero. Hence the determinant in (\ref{normB}) resembles very much (\ref{Norm0}). There is however a different prefactor which arises because (\ref{normB}) was computed using the Algebraic Bethe Ansatz normalization whereas (\ref{Norm0}) was obtained in the coordinate Bethe Ansatz formalism. Of course, in (\ref{main}), the normalizations of the kets are irrelevant. However, to compare the intermediate quantities (\ref{normB}) and (\ref{Norm0}) we better know the relation between the two. When there are no impurities the relative normalization is (\ref{lambda}). Taking it into account we see that (\ref{normB}) and (\ref{Norm0}) agree in the homogeneous limit $\theta_j\to 0$.
\begin{figure}[t]
\begin{center}
\includegraphics[trim=0cm 9cm 0cm 0cm, clip=true, scale=.6]{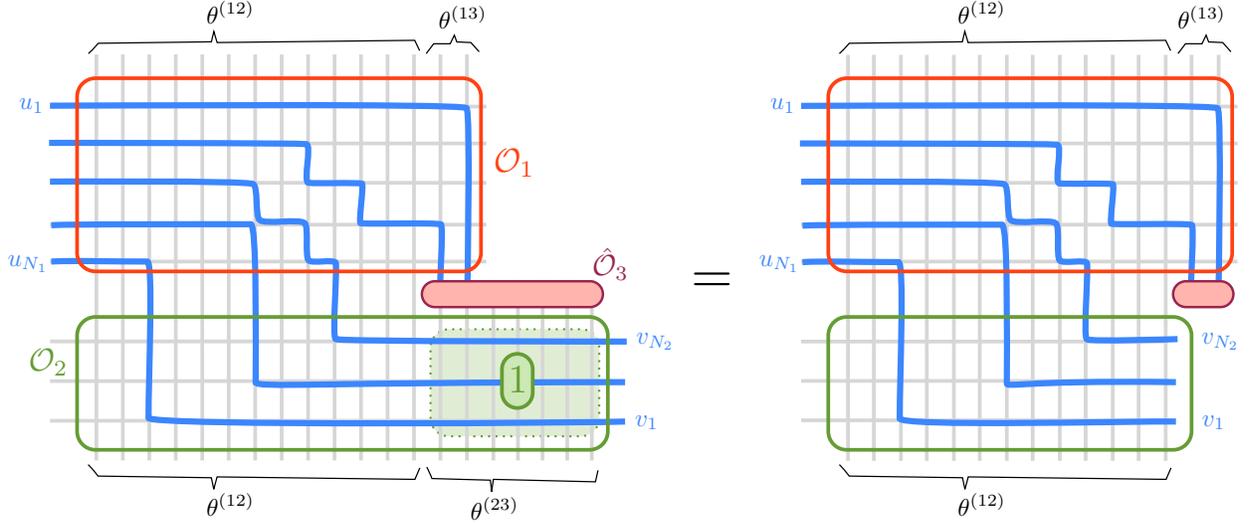}
\end{center}
\caption{Main scalar product appearing in the impure structure constants (\ref{omar}). When contracting the three operators the rightmost sites of the spin chain $\O_2$ are contracted with empty sites in the chain of the third operator. Hence the weight associated with the lower right corner in the right picture is just $1$ since the weight for each vertex where a blue thick lines crosses a thin gray line is $1$, see (\ref{stateAlg}). Because of this, this scalar product does not depend on the impurities $\theta^{(23)}$ at all and can be reduced to the object in the right \cite{omar}.
} \la{figC123}
\end{figure}

Finally we have the numerator which can be represented graphically as in figure \ref{figC123}. We have \cite{omar}
\beq
\<\theta^{(1)} ;{\bf u} | \hat{\mathcal{O}}_3| \theta^{(2)};{\bf v} \>=\frac{\displaystyle
\prod_{m}^{N_3}\prod_{n}^{N_1}(u_n- \hat\theta^{(1)}_{m}+i/2)/\prod_{m}^{N_3}\prod_{n}^{N_2}(v_n- \hat\theta^{(1)}_{m}+i/2)
}{\displaystyle
\prod_{n<m}^{N_1}(u_m-u_n)
\prod_{n<m}^{N_2}(v_n-v_m)
\prod_{n<m}^{N_3}( \hat\theta^{(1)}_{n}-\hat\theta^{(1)}_{m}) }
\det\Big([G_{nm}] \oplus [F_{nm}] \Big) \la{omar}
\eeq where $\hat \theta^{(1)}_m=\theta^{(1)}_{L_1+1-m}$ and
\beq
F_{nm}=\frac{1}{(u_{n}-{\color{black}\hat\theta}_m^{(1)})^2+\frac{1}{4}} \,, \qquad G_{nm}=
\prod_{a=1}^{L_1} \frac{v_m-\theta_a^{(1)}+i/2}{v_m-\theta_a^{(1)}-i/2}
\frac{\prod_{k\neq{   n}}^{N_1}(u_k-v_{   m}\!+\! i)}{u_n-v_m}-
\frac{\prod_{k\neq{   n}}^{N_1}(u_k-v_{   m}\!-\! i)}{u_{   n}-v_{   m}} \nn\;.
\eeq
Where $F_{nm}$ is $N_1\times N_3$ and $G_{nm}$ is $N_1\times N_2$ matrices
completing together to a $N_1\times N_1$ square matrix.  Note that only the $\theta^{(1)}=\theta^{(12)}\cup\theta^{(13)}$ appears in the numerator factor. There is no dependence on the impurities $\theta^{(23)}$ associated to lines going from $\O_2$ to $\O_3$. From the combinatorial point of view this is obvious, see caption of figure \ref{figC123}.

The homogeneous limit is a bit more tricky here, see \cite{omar} for details.
One has to resolve a $0/0$ uncertainty using standard l'H\^opital's rule.
On the one hand the prefactor in (\ref{omar}) diverges as $\theta_j\to 0$. On the other hand the determinant vanishes since $F_{nm}$ will no longer depend on the column index $m$ as $\theta_m\to 0$.\footnote{Once we take the limit carefully the result is proportional to  the determinant
\beqa
\left|\begin{array}{cccccc}
\frac{\partial T(v_1)}{\partial u_1} & \dots &  \frac{\partial T(v_{N_2})}{\partial u_1} & q_2(u_1) & \dots & q_{N_3+1}(u_1)\\
\vdots & & \vdots & \vdots & & \vdots \\
\frac{\partial T(v_1)}{\partial u_{N_1}} & \dots &  \frac{\partial T(v_{N_2})}{\partial u_{N_1}} & q_2(u_{N_1}) & \dots & q_{N_3+1}(u_{N_1})
\end{array}
\right| \,\,, \qquad \begin{array}{l}
q_n(u)\equiv i(u+i/2)^{-n+1}-i(u+i/2)^{-n+1}\\
T(u)\equiv \frac{Q(u-i)}{Q(u)} + \(\frac{u-i/2}{u+i/2}\)^{L_1} \frac{Q(u+i)}{Q(u)}\\
Q(u) \equiv \prod_{j=1}^{N_1} (u-u_j)
\end{array}
\la{theDet}
\eeqa
The expression (\ref{theDet}) is the most complicated factor in the tree-level structure constant for three non-BPS operators. In section (\ref{coorSec}) we tried to guess the higher loop generalization of several simpler scalar products rather successfully. Guessing the generalization of (\ref{theDet}) turns out to be quite non-trivial as mentioned before.
}

In what follows we will also need a particular case of the general scalar product (\ref{main}).
That is the case when ${\cal O}_1$ degenerates to BPS.
This can be done by sending all roots $u_j \to \infty$. In this limit
 the state $|1\>$ becomes a vacuum descendent and (\ref{omar}) reduces to the generalization of the DVD amplitude with impurities,
\beq
\left\<\,\ssststile{\,\,\,\,L_1-N_3\,\,\,\,}{\,\,N_1-N_3\,\,}\,\,\ssststile{\,\,\,\,L_3-N_3\,\,\,\,}{\,\,0\,\,}\,\right|
 \theta^{(2)};{\bf v}\Big \>=\sum_{\alpha \cup \bar \alpha = {\bf v}} (-1)^{|\bar \alpha|} \[\prod_{\bar a \in \bar\alpha}  \prod_{n=1}^{L_{12}} \frac{\bar a-\theta_n+i/2}{\bar a-\theta_n-i/2}\] \prod_{a\in \alpha\,, \bar a\in \bar \alpha} f(a,\bar a)
\eeq
where $f(u,v)=1+i/(u-v)$. Denoting the square bracket by $e_{\bar \alpha}^{L_{12}}$ and using the same kind of short-hand notation from \cite{paper1} we can compress this expression to
\beq\la{tree132}
\<\text{vac}|
 \theta^{(2)};{\bf v}\> =
\sum_{\alpha \cup \bar \alpha = {\bf v}}
(-1)^{|\bar\alpha|}
f^{\alpha,\bar\alpha}e_{\bar \alpha}^{L_{12}}\,,\,\,\, \text{where} \qquad \<\text{vac}|=\left\<\,\ssststile{\,\,\,\,L_1-N_3\,\,\,\,}{\,\,N_1-N_3\,\,}\,\,\ssststile{\,\,\,\,L_3-N_3\,\,\,\,}{\,\,0\,\,}\,\right|\;.
\eeq
Equation (\ref{main}) -- and also equations (\ref{normB}) and (\ref{tree132}) -- will be used in the next section.
As we will see later, by use of well designed differential operators we will be able to promote  tree level
``impure" expressions to higher loops ``pure" results. Developing the formalism to do so will be the subject of the next section.

\section{$\Theta$--morphism and Quantum Corrected States} \la{secMor1}
In this section we introduce the main object of this paper: the $\Theta$--morphism.
\subsection{One Loop $\Theta$--Morphim. Definition and Properties}

The $\Theta$--morphism is a coupling dependent functional acting on functions of a set of variables $\theta\equiv \{\theta_i\}_{i=1}^L$
which we denote by double brackets
\beq
\Theta\;:\;f(\theta)\;\mapsto\;\ckappa{\theta}{}{f(\theta)}\;.
\eeq

In section \ref{Axiomatics} we will explain how one can define this operator by requiring it to obey several defining axioms.
The justification of the name will be clear from what follows. At leading order the $\Theta$--morphim is given by the differential operator
\beq
\ckappa{\theta}{}{f(\theta)}\equiv  \left. f+\frac{ g^2}{2}\sum_{i=1}^{L} \mathcal{D}_i^2 f + \mathcal{O}(g^4) \right|_{\theta_j \to 0}\;. \la{mor1}
\eeq
where
\beq
\mathcal{D}_i = \partial_i- \partial_{i+1} \qquad \text{and} \qquad  \partial_i \equiv \frac{\partial}{\partial \theta_i}\,.
\eeq
we use a periodic boundary conditions so that $\mathcal{D}_L=\partial_L-\partial_1$.
For now let us take the definition (\ref{mor1}) as our starting point. For further motivation see the next subsections, \cite{short} and \cite{GSV}.
In particular, we will see below that by acting with this operator on the impure states of section \ref{impure}, we generate perfect two loop eigenvectors of the Dilatation operator (\ref{dilatation0}) including all the messy contact terms discussed in section \ref{contactSec} and appendix \ref{C4}! This is of course the main justification for (\ref{mor1}).

For now, let us explore some properties of the map (\ref{mor1}) that will come in handy later on. Acting on a product of two functions we have
\beq
\ckappa{\theta}{}{f(\theta)h(\theta)}=
\ckappa{\theta}{}{f(\theta)}
\ckappa{\theta}{}{h(\theta)}+
\cckappa{\theta}{}{f(\theta)}{h(\theta)}
\;. \la{mor0}
\eeq
where
\beq
\cckappa{\theta}{}{f}{h}\equiv g^2\sum_{i=1}^{L} \mathcal{D}_i f \,\mathcal{D}_i h\,. \la{ct}
\eeq
In the right hand side we  set to zero all impurities after taking the derivatives, just like we did in (\ref{mor1}). In what follows this is always implicit; we always set $\theta_j\to 0$ at the end. We denote the contribution (\ref{ct}) by \textit{cross-term}. If it were not for this term then (\ref{mor0}) would be the mathematical definition of a morphism
\footnote{We are grateful to B.~Vicedo for proposing this name to us}
. Hence, the cross-term is the deviation from the morphism property.

The cross-term (\ref{ct}) does vanish if one of the functions is a totally symmetric function of the $\theta_i$. That is we do have the restricted morphism relation
\beq
{\bf Morphism:}\;\;\;\;\ckappa{\theta}{}{f_{\rm sym}(\theta)h(\theta)}=
\ckappa{\theta}{}{f_{\rm sym}(\theta)}
\ckappa{\theta}{}{h(\theta)}\;. \la{morpro}
\eeq
Another nice property that will be used shortly is that
\beq
{\bf Dispertion:}\;\;\;\;\ckappa{\theta}{}{\sum_{i=1}^L\log\frac{u-\theta_i+\tfrac i2}{u-\theta_i-\tfrac i2}}=
L\log\frac{x(u+\tfrac i2)}{x(u-\tfrac i2)} + \mathcal{O}(g^4)\;\, , \la{disppro}
\eeq
which one can easilly check directly using (\ref{mor1}). In this expression $x(u)$ is the so called Zhukowsky variable
\beq
x(u)+\frac{1}{x(u)} \equiv \frac{u}{g} \,, \qquad x(u) = \frac{u}{g}-\frac{g}{u}+\mathcal{O}(g^3) \,. \nn
\eeq
These relations are very important and have  interesting implications. 
 For example, an obvious consequence of the morphism relation (\ref{morpro}) is that for any function $F$ we have
\beq
\ckappa{\theta}{}{F \Big[ f_{\rm sym}(\theta) \Big]}=
F\[\ckappa{\theta}{}{f_{\rm sym}(\theta)}\] \,. \la{lema}
\eeq
Another nice implication follows when we combine both properties:  Suppose we are dealing -- as we often will -- with a function $f( { \bf u}, \theta)$ which is also a function of
 the Bethe roots ${\bf u} = \{u_1,\dots, u_N\}$. These are quantized according to Bethe equations (\ref{BAEimpure}),
\beq
\sum_{a=1}^{L} \frac{1}{i} \log\(\frac{u_k-\theta_a+i/2}{u_k-\theta_a-i/2}\)+
\sum_{m\neq k}^{N} \frac{1}{i}\log\( \frac{u_k-u_m-i}{u_k-u_m+i}\) = 2\pi n_k \,, \qquad n_k \in \mathbb{Z} \la{notanymore}
\eeq
In particular, they depend on the impurities $\theta_j$ in a totally symmetric way.
Then, a consequence of the above relations is that $\ckappa{\theta}{}{f( { \bf u}, \theta)}$ is still given by (\ref{mor1}) where
\begin{enumerate}
\item we can treat the $u_j$ as independent of $\theta_j$ when taking the derivatives in (\ref{mor1}) \textit{but}
\item after applying the morphism -- i.e. in the right hand side of (\ref{mor1}) -- the Bethe roots are now quantized according to the BDS equations \cite{BDS}
\beq
\phi_k^{\bf u} \equiv \frac{L}{i} \log\frac{x(u_k+i/2)}{x(u_k-i/2)}+
\sum_{m\neq k}^{N} \frac{1}{i}\log\( \frac{u_k-u_m-i}{u_k-u_m+i}\) = 2\pi n_k \,, \qquad n_k \in \mathbb{Z} \la{BAEimpure2}
\eeq
instead of (\ref{notanymore}). The equations (\ref{BAEimpure2}) are nothing but the corrected Bethe equations (\ref{BetheEq0}) and (\ref{Smatrix}). \end{enumerate}
Of course, such a prescription is exactly what one would like when dealing with expressions involving Bethe roots!

\subsection{Single Derivative of Monodromy Matrix et al} \la{IdSec}
The most important object in the Algebraic Bethe ansatz construction is the monodromy matrix
\beq
\hat L(\theta;u) \equiv  \(\mathbb{I}_{10} +\frac{i}{u-\theta_1-i/2} \mathbb{P}_{10} \) \otimes \dots \otimes \(\mathbb{I}_{L0} +\frac{i}{u-\theta_L-i/2} \mathbb{P}_{L0} \) \la{Ldef}
\eeq
Operators with subscript $ij$ act on $V_i \otimes V_j$ where $V_{1},\dots,V_L$ are the Hilbert spaces at each spin chain site while $V_0$ is an auxiliary space which is also $\mathbb{C}^2$. That is, we can think of $\hat L(u)$ as a $2\times 2$ matrix whose entries are operators acting on the full physical Hilbert space $\mathcal{H}=V_1\otimes \dots \otimes V_L$,
\beq
\hat L(\theta;u) =\( \begin{array}{cc}
\hat A(u) & \hat B(u) \\
\hat C(u) & \hat D(u) \end{array} \)
\eeq
The operators $\hat B(u)$ are creation operators; the Bethe eigenstates are simply given by (\ref{BBOmega}). Graphically, this action can be represented as in (\ref{stateAlg}). A very important relation that we will use extensively is then
\beq
i{\cal D}_j\hat L=-
[\mathbb{H}_{j,j+1}, \hat L]+\delta_{j,L}\sum_{n=1}^L[ \mathbb{H}_{n,n+1}, \hat L] \,. \la{basic}
\eeq
where $\mathbb{H}_{j,j+1}=\mathbb{I}_{j,j+1} - \mathbb{P}_{j,j+1}$. As usual, we have set to zero all impurities after taking the derivatives. Hence, when acting on Bethe states $|\theta;{\bf u}\> = \hat B(u_1) \dots \hat B(u_N) \left| \uparrow \dots \uparrow \right\> $ with a single derivative $\mathcal{D}_j$ we get from (\ref{basic})  that $i{\cal D}_j |\theta;{\bf u}\> = - \mathbb{H}_{j,j+1}|0;{\bf u}\> + \delta_{j,L}\sum_{n=1}^L \mathbb{H}_{n,n+1} |0;{\bf u}\>  $
and since they are Bethe eigenstates the last term can be replaced by the state energy so that
\beq
i{\cal D}_j |\theta;{\bf u}\> = - \mathbb{H}_{j,j+1} |0;{\bf u}\> + \delta_{j,L} \Gamma_{\bf u} |0;{\bf u}\>  \,. \la{PsiId}
\eeq
where
\beq
\Gamma_{\bf u} = \sum_{j=1}^N \frac{1}{u_j^2+\frac{1}{4}} \,.
\eeq
Another very important object is the transfer matrix
\beq
\hat T(\theta;u) \equiv \Tr\, \hat L(\theta;u) = \hat A(u) + \hat D(u) \,. \la{Tdef}
\eeq
To the order we will be working at -- i.e. as an expansion in small $\theta_j$'s -- this object commutes with the Hamiltonian $\sum_{n=1}^L \mathbb{H}_{n,n+1}$. Hence, from (\ref{basic}) we find
\beq
i{\cal D}_j\hat T=-
[\mathbb{H}_{j,j+1}, \hat T] \,. \la{Tid}
\eeq
The identities (\ref{basic}), (\ref{PsiId}) and (\ref{Tid}) will be extensively used in what follows.

\subsection{Integrability and Two Loop Eigenstates} \la{algStatesSec}
Consider the impure states (\ref{BBOmega}).
They diagonalize the transfer matrix (\ref{Tdef}),\footnote{The eigenvalue reads \beq
{T}(\theta;v)=a(\theta;v)\frac{Q_{\bf u}(v-i)}{Q_{\bf u}(v)}+\frac{Q_{\bf u}(v+i)}{Q_{\bf u}(v)} \,, \text{ where} \,\,\,\,\,
a(\theta;v) \equiv \prod_{a=1}^L \frac{v-\theta_a+i/2}{v-\theta_a-i/2} \,, \,\,\,\,\, Q_{\bf u} (v) \equiv \prod_{j=1}^N (v-u_j) \la{footeq}
\eeq
}
\beq
{\hat T}(\theta;v)|\theta;{\bf u}\rangle={T}(\theta;v)|\theta;{\bf u}\rangle \la{diag} \,.
\eeq
The eigenvalue ${T}(\theta;v)$ is a symmetric function of the $\theta_j$'s.
A most striking fact about (\ref{diag}) is that the states $|\theta;{\bf u}\rangle $ do \textit{not} depend on the spectral parameter $v$ and still they diagonalize the full family of operators ${\hat T}(\theta;v)$ for \textit{any} value of $v$! Integrability, that is exact solvability of the model, lies upon this remarkable fact. In this section we want to understand what happens to such remarkable properties  once we apply the $\Theta$--morphism to relations like (\ref{diag}).

Let $\hat F(v)$ be an arbitrary function of
${\hat T}(\theta;v)$ and $F(v)$ its eigenvalue. Then
\beq
\hat F(v) |\theta;{\bf u}\rangle=F(v) |\theta;{\bf u}\rangle \,.
\eeq
Applying the $\Theta$--morphism we get
\beq
\ckappa{\theta}{}{\hat F}
\ckappa{\theta}{}{ |\theta;{\bf u}\rangle}+
\cckappa{\theta}{}{\hat F}{ |\theta;{\bf u}\rangle}
=
\ckappa{\theta}{}{F}
\ckappa{\theta}{}{|\theta;{\bf u}\rangle}\la{stepone} \,.
\eeq
in the right hand side we get no cross term since $F$ is a symmetric function of $\theta$'s, see (\ref{morpro}).
Consider the cross-term in the left-hand side. Using the relations derived in the previous section we have
\beq
\cckappa{\theta}{}{\hat F}{ |\theta;{\bf u}\rangle}=-g^2\sum_{j=1}^L [P_{j,j+1},{\hat F}]P_{j,j+1}
|0;{\bf u}\rangle
+g^2[{\mathbb H}_{L,1},{\hat F}]\Gamma_{\bf u}
|0;{\bf u}\rangle \la{step2}
\eeq
As usual we set the impurities to zero at the end in the right hand side. Note that at this order -- since both terms are already multiplied by a $g^2$ factor -- we can replace either of the kets $|0;{\bf u}\rangle$ by $\ckappa{\theta}{}{|\theta;{\bf u}\rangle}$ if we want to.
Note also that only the last term in (\ref{step2}) does not have have nice cyclicity properties; this term can be interpreted as a correction to the state.
More precisely, if we define
\beq\la{closed}
|{\bf u}\rangle\equiv
(1-g^2 \Gamma_u {\mathbb H}_{L,1})
\ckappa{\theta}{}{|\theta;{\bf u}\rangle}
\eeq
then  (\ref{stepone}) becomes
\beq
\hat{\mathcal{F}}(v) |{{\bf u}}\rangle =
{\mathcal{F}}(v) |{{\bf u}}\rangle \la{after}
 \eeq
where
\beq
\hat{\mathcal{F}} \equiv \ckappa{\theta}{}{\hat F}
-g^2\sum_{j=1}^L [P_{j,j+1},{\hat F}]P_{j,j+1}
\qquad \text{and}\qquad \mathcal{F} \equiv
 \ckappa{\theta}{}{F} \la{calF}\;.
\eeq
The very exciting property we highlighted above is present again in (\ref{after}). Namely, we have a set of states $| {{\bf u}}\>$ which do not depend on the spectral parameter and still diagonalize a full family of operators $\hat{\mathcal F}(v)$ for arbitrary $v$. In this sense, we have integrability before and after the $\Theta$--morphism.

\subsection{Two-loop Hamiltonian}
We found the first sign of evidence in favor of the $\Theta$--morphism in (\ref{BAEimpure2}). In these Bethe equations we identified the corrected momentum
\beq
p(u) = \frac{1}{i} \log \frac{x(u+i/2)}{x(u-i/2)}\;.
\eeq
Now we should look for the corrected energy
\beq
\epsilon(u) = \frac{2ig}{x(u+i/2)} -\frac{2ig}{x(u-i/2)} 
\eeq
More precisely we want to find some operator $\hat F$ such that $\hat{\mathcal{F}}$ defined through (\ref{calF}) leads to the two loop Dilatation operator (\ref{dilatation0}). We will look for this operator by fixing the eigenvalue $\mathcal{F}$ to be equal to the energy
\beq
\gamma=\sum_{j=1}^L\epsilon(u_j) =
\sum_{j=1}^N \[
\frac{2ig^2}{u_j+\tfrac i2}
+
\frac{2ig^4}{(u_j+\tfrac i2)^3}+c.c. \] + \mathcal{O}(g^6) \,.\la{energyc}
\eeq
Using (\ref{footeq}) we can directly check that\footnote{We would still get (\ref{energyc}) without the $\Theta$--morphism $\ckappa{\theta}{}{}$, that is if we simply set the impurities to zero and not take any derivatives. Of course, in that case the rapidities would obey (\ref{BAEimpure}) with $\theta_j=0$, that is they would be solutions to the one loop Bethe equations which is not what we want. If we apply the $\Theta$--morphism we do get  the two loop corrected Bethe equations (\ref{BAEimpure2}) as desired. Indeed, as we can see directly from (\ref{three}) below, we would not get the correct two loop Hamiltonian without the morphism contribution.}
\beqa
\gamma=\left.\ckappa{\theta}{}{2ig^2\d_u\log T+ig^4\d_u^3\log T}\right|_{u=-i/2} \, \la{gammaT} \,.
\eeqa
Hence, following (\ref{calF}), the candidate operator that should correspond to the two loop Dilatation operator is\footnote{At this loop order $g^4 i\d_u^3\log \hat T$ and $g^4 i\d_u^3 \ckappa{\theta}{}{\log \hat T}$ are indistinguishable.}
\beq
\left.\ckappa{\theta}{}{2g^2i\d_u\log \hat T}+g^4 i\d_u^3\log \hat T
-2 g^4\sum_{j=1}^L [P_{j,j+1},i\d_u\log \hat T]P_{j,j+1}
\right|_{u=-i/2} \la{todo}\;.
\eeq
Using the definition (\ref{Tdef}) and (\ref{Ldef}) we find indeed that these three contributions lead to
\beqa
&&\left\{\begin{array}{c}
\ckappa{\theta}{}{2g^2i\d_u\log \hat T} \\
g^4 i\d_u^3\log \hat T \\
-2 g^4\sum_{j=1}^L [P_{j,j+1},i\d_u\log \hat T]P_{j,j+1} \end{array}\right\}_{u=-i/2} = \left\{\begin{array}{c}
2g^2-8g^4 \\
-\,4g^4\\
+\,4g^4 \end{array}\right\} \sum_{i=1}^L (\mathbb{I}-\mathbb{P}_{i,i+1})+ \nn\\
&&\qquad+ \left\{\begin{array}{c}
+\,4g^4 \\
+\,2g^4\\
-\,4g^4 \end{array}\right\} \sum_{i=1}^L (\mathbb{I}-\mathbb{P}_{i,i+2})+ \left\{\begin{array}{c}
-\,2g^4 \\
+\,2g^4\\
-\, 0\end{array}\right\}\sum_{i=1}^L \[ {\mathbb P}_{i,i+1} , \[ {\mathbb P}_{i+1,i+2}, {\mathbb P}_{i+2,i+3}\]\]  \la{three}
\eeqa
Adding the three quantities we see that (\ref{todo}) is exactly equal to
\beq
\mathbb{H}=(2g^2-8g^4)\sum_{i=1}^L (\mathbb{I}-\mathbb{P}_{i,i+1})+2g^4\sum_{i=1}^L (\mathbb{I}-\mathbb{P}_{i,i+2})
\eeq
which is the Dilatation operator of $\mathcal{N}=4$ SYM up to two loops \cite{BDS}, see (\ref{dilatation0}). Let us summarize the outcome of these last sections. We proved that

\textit{The states (\ref{closed}), (\ref{BBOmega}) are the precise two loop eigenvectors of the two loop Dilatation operator (\ref{dilatation0}) of $\mathcal{N}=4$ SYM. In particular, all the monstrous contact terms discussed in section \ref{contactSec} and appendix \ref{C4} are automatically incorporated in the algebraic approach (\ref{closed}) using the $\Theta$--morphism!}

This is one of the main results of this paper.  Higher loops will be briefly touched upon in section \ref{Axiomatics}. In section \ref{ThetaC123} we will apply these results to the computation of structure constants at one loop level.

\subsection{$\Theta$--morphism and the All Loop Dispersion} \la{Axiomatics}
In this section we will briefly discuss the $\Theta$--morphism from a more axiomatic point of view. The goal is to find out how we can constrain the $\Theta$--morphism action at higher loops. We start by an ansatz of the form
\beq
\ckappa{\theta}{}{f(\theta)}=
\left.
\exp\(\sum_{i=1}^L\sum_{l=1}^{\infty} g^{2l} P_{2l}(\d_i,\dots,\d_{i+l})\)f(\theta)\right |_{\theta_i=0} \la{locality}
\eeq
where $P_{2l}$ is polynomial of degree $2l$.
In what follows we assume that $P_{2l}$ is a homogeneous polynomial of degree $2l$.
Note that in the exponent the polynomial only acts on $l$ neighboring sites. We call this axiom the \textit{locality axiom}. It is inspired by the structure of perturbation theory where the interaction range also increases with the perturbative loop order.

Next we require the second axiom:
\beq
\ckappa{\theta}{}{f_{\rm sym}(\theta)h(\theta)}=
\ckappa{\theta}{}{f_{\rm sym}(\theta)}
\ckappa{\theta}{}{h(\theta)}\;. \la{morpro2}
\eeq
which we call the \textit{morphism axiom}. As explained before, this is a very important property which allows us to treat the Bethe roots as numbers, ignoring their dependence on the impurities $\theta_j$. This seems like a very mild requirement but it turns out to be extremely powerful! We can study the constraint (\ref{morpro2}) in perturbation theory. We did it up to order $g^8$ by considering polynomial functions of increasingly higher degree. The result of this brute force analysis is quite surprising. We found that the morphism axiom fixes completely\footnote{The morphism axiom is by no means sufficient to constrain (\ref{locality}) completely. It does not fix the form of the polynomials $P_{2l}$ at higher $l$. However, it does seem to fix the parameters of the polynomials that enter in the computations of $\ckappa{\theta}{}{\theta_j^n}$. This is quite nice because it is exactly what we need to act on the dispersion relation!}
\beqa
\ckappa{\theta}{}{\theta_j^4} &=& 3 \,\,\,\,g^2\ckappa{\theta}{}{\theta_j^2}  = \,3 \,\,\,g^4 \nn\\
\ckappa{\theta}{}{\theta_j^6} &=& 10  \,g^{4}\,\ckappa{\theta}{}{\theta_j^2}=10  \,g^{6}  \nn\\
\ckappa{\theta}{}{\theta_j^8} &=& 35\, g^{6}\, \ckappa{\theta}{}{\theta_j^2} =35 \, g^8\la{seq}
\eeqa
The action on $\theta_j^2$ is not constrained and amounts to (re)defining the coupling constant so we set it to $g^2$. We can ask the famous
\textit{The On-Line Encyclopedia of Integer Sequences}\footnote{http://oeis.org/ or directly http://oeis.org/search?q=3\%2C+10\%2C+35}
\beq
\ckappa{\theta}{}{\theta_j^{2n}}= \frac{(2 n)!}{2 (n!)^2}g^n \, \la{goal}
\eeq
which are very familiar in this context \cite{BDS,didina}.
We conjecture that (\ref{locality}) and (\ref{morpro2}) imply (\ref{goal}). We could not find a prove of this simple statement. If true, (\ref{goal}) immediately implies something quite remarkable: \textit{At all loop orders}
\beq
\ckappa{\theta}{}{\log\prod_{i=1}^L\frac{u-\theta_i+\tfrac i2}{u-\theta_i-\tfrac i2}}=
L\log\frac{x(u+\tfrac i2)}{x(u-\tfrac i2)}\;
\eeq
where $x(u)$ are the Zhukowsky variables.
We see that the one loop momentum (with impurities) becomes promoted to the all loop momentum! We believe this is a very strong evidence in favor of this approach. We will continue our first principles study of the $\Theta$--morphism elsewhere \cite{GSV}.

\section{$\Theta$--morphism and One Loop Structure Constants} \la{ThetaC123}

In this section we will use the $\Theta$--morphism to derive the one loop structure constants by  promoting the tree level impure structure constants of section \ref{impure}. Section \ref{coorSec} was also devoted to the study of one loop structure constants at one loop. The main differences between the approach in this section and the method used in section \ref{coorSec} are
\begin{itemize}
\item First, here we will use the Algebraic Bethe ansatz approach together with the $\Theta$--morphism whereas in section \ref{coorSec} we followed the Coordinate Bethe ansatz. Of course, at the end of the day, both formalisms will give exactly the same result for physical quantities such as  (the absolute value of) structure constants!

However, in intermediate stages, scalar products in the coordinate and algebraic basis will differ since the kets are  normalized differently. It is convenient to know the relative normalization to be able to compare not only the final results but also the intermediate ones. For example, at one loop we have (\ref{lambda}). By checking a few examples like (\ref{2ptExample}) at two loops, we can find the relative normalization between the two loop coordinate states -- which we denoted by $|{\bf p}\>$ in section \ref{contactSec} -- and the algebraic states  -- which we denoted by $|{\bf u}\>$ in section \ref{algStatesSec}. We find
\beq
|{\bf u}\> = \mu \,|{\bf p}\> \,, \qquad \mu=\(1- \frac{\gamma({\bf p})^2}{4g^2} \) \prod_{j=1}^N(e^{-ip_j}-1) \prod_{i<j} {\frak f}(p_i,p_j) \la{conv1}
\eeq
where (\ref{gamma0}) and (\ref{fcorr}).

Another minor difference is that in the Coordinate Bethe ansatz (sections \ref{contactSec} and \ref{coorSec}) we use mostly the momenta $p_a$ as variables. In the Algebraic Bethe ansatz (sections \ref{algSec} and \ref{secMor1}) we use mostly the Bethe rapidities $u_a$ as our variables. The relation between the two is given by $u_a=u(p_a)$ where (\ref{Smatrix}).
Equivalently
\beq
e^{ip} = \frac{x(u+i/2)}{x(u-i/2)} \,, \qquad x(u)= \frac{u+\sqrt{u^2-4g^2}}{2g} = \frac{u}{g}-\frac{g}{u} + \dots
\eeq
We will also use $x^{\pm}=x(u\pm i/2)$ and $x_a^{\pm}=x(u_a\pm i/2)$.  For example, the above equation (\ref{conv1}) can equivalently be written as
\beq
\mu=({1- g^2\Gamma_u^2}) \prod_{j=1}^N\(\frac{x_j^-}{x_j^+}-1\) \prod_{i<j} {\frak f}(u_i,u_j) \la{conv2}
\eeq
where
\beq
\Gamma_{\bf u} \equiv \sum_{j=1}^N \frac{1}{u_j^2+\frac{1}{4}}\,, \qquad  {\frak f}(u,v)\equiv \(1+\frac{i}{u-v}\)\(1+\frac{g^2}{(u^2+\frac{1}{4})(v^2+\frac{1}{4})}\)
\eeq
The set of momenta of the three operators $\O_1$, $\O_2$ and $\O_3$ is, respectively ${\bf k}$, ${\bf p}$ and ${\bf q}$ while the corresponding rapidities are denoted by ${\bf u}$, $\bf v$ and $\bf w$.

\item The second important difference between the two approaches is in the method used.
In the coordinate approach of section \ref{coorSec} we start with the scalar products for the one loop eigenstates and we \textit{guessed} the higher loop generalizations. These conjectures were based on (1) playing with examples with a small number of excitations and finding patterns and (2) using the experience coming from higher loop corrections in the spectrum problem. 
Sometimes, one can guess not only the first quantum correction but even conjecture all loop results, see e.g. section \ref{normsec}. However, for the most complicated expectation value \verb"involved" (\ref{tocompute}) the guessing strategy was manifestly insufficient  to guess even the first quantum correction. With the algebraic Bethe ansatz approach we have full control over the two loop eigenvectors. As such, all one loop structure constants can be derived using the $\Theta$--morphism without any guessing.

In sum, the two methods are nicely complementary.
\end{itemize}

\subsection{Derivation of the Norm}
In (\ref{tocompute}) we identified all required scalar products and expectation values that we need to compute to obtain the complete one loop corrected structure constants in the $SU(2)$ setup of \cite{paper1}. We will now compute them using the algebraic Bethe ansatz approach.

In this section we consider the simplest ones, the norms $\<{\bf 1}|{\bf 1}\>$ etc. The norms were studied in the coordinate Bethe ansatz language in section \ref{normsec}. In the algebraic Bethe ansatz the corrected two loop eigenvectors are given by (\ref{closed}), $|{\bf 1}\> = |{\bf u}\>$ so what
we want to compute is
\beq\la{closednorm}
\langle{\bf u}|{\bf u}\rangle\equiv
\ckappa{\theta}{}{\langle\theta;{\bf u}|}
(1-2g^2 \Gamma_u {\mathbb H}_{L,1})
\ckappa{\theta}{}{|\theta;{\bf u}\rangle}\;.
\eeq
Note that for cyclic states the insertion of the operator ${\mathbb H}$
will not depend of the position where it is inserted.
This observation allows for the replacement ${\mathbb H}_{L,1}\to \Gamma_{\bf u}/L$ since we can average over where we insert. We will not need to use this in the derivation.

The strategy is now to convert the contraction of two $\Theta$-deformed
states to the overall $\Theta$-deformation of the scalar product
\beq\la{Ntheta}
\ckappa{\theta}{}{\langle\theta;{\bf u}|\theta;{\bf u}\rangle}\;.
\eeq
Note that we do know the explicit expression for the quantity inside the brackets; it is given in (\ref{normB}). Hence evaluating (\ref{Ntheta}) is straightforward (we will do it below).

To bring (\ref{Ntheta}) to a form closer to (\ref{closednorm}) we have to compute the cross-term. Using the identities derived in section \ref{IdSec}, in particular (\ref{PsiId}), we easily find
\beqa
\cckappa{\theta}{}{\langle\theta;{\bf u}|}{|\theta;{\bf u}\rangle}
&=&\sum_{j=1}^L g^2\langle 0;{\bf u}|({\mathbb H}_{j,j+1}-\delta_{j,L}\Gamma_{\bf u})^2|0;{\bf u}\rangle\\
&=&g^2\langle 0;{\bf u}|2\Gamma_{\bf u}+\Gamma_{\bf u}^2-2\Gamma_{\bf u}{\mathbb H}_{1,L}|0;{\bf u}\rangle\;. \la{this}
\eeqa
We see that the term with insertion of the Hamiltonian density is exactly the same as in (\ref{closednorm})!\footnote{In (\ref{this}) we can replace $|0;{\bf u}\rangle$ by $\ckappa{\theta}{}{|\theta;{\bf u}\rangle}$ if we want to. At this loop order they are indistinguishable since (\ref{this}) already has an overall $g^2$ factor.}
Hence
\beq
\langle{\bf u}|{\bf u}\rangle =(1-2g^2 \Gamma_{\bf u}-g^2 \Gamma_{\bf u}^2) \ckappa{\theta}{}{\langle\theta;{\bf u}|\theta;{\bf u}\rangle}\;, \la{normPart1}
\eeq
which is exactly the kind of expression we were after. Furthermore, as mentioned above, (\ref{Ntheta}) can be readily computed: the function (\ref{normB}) only depends on the impurities $\theta_j$ through the symmetric combination appearing in (\ref{disppro}). Hence, using (\ref{disppro}) and (\ref{lema}) we immediately conclude that all we need to do is replace the dispersion factor, that is
\beq
 \ckappa{\theta}{}{\langle\theta;{\bf u}|\theta;{\bf u}\rangle}=\prod_{m\neq k} \frac{u_k-u_m+i}{u_k-u_m}   \det_{j,k\le N_1} \frac{\partial }{\partial u_j} \[ \frac{L}{i}  \log \frac{x(u_k+i/2)}{x(u_k-i/2)}
+\frac{1}{i}\sum_{m\neq k}^{N_1} \log\frac{u_k-u_m-i}{u_k-u_m+i}\]   \nn
\eeq
Taking into account the conversion factor (\ref{conv1}) one can check that this is in perfect agreement with the conjecture for the norm of section \ref{normsec}!

\subsection{Derivation of the DVD Amplitude}
Having computed the norms we now move to the next line in (\ref{tocompute}), the contribution \verb"simple". As explained in section \ref{simplificationSec} this quantity can be simplified further so that all we need is the scalar product (\ref{Adef}) between a Bethe state and a state obtained by gluing together two different vacuum descendents.

For one loop eigenvectors with impurities (\ref{BBOmega}) this quantity is given in  (\ref{tree132}).
On the other hand, we are interested in the two loop eigenvectors with no impurities (\ref{closed}), that is
$
\<\text{vac}| {\bf v} \>
$.
To relate the two we follow the same strategy as in the previous section and start by applying the $\Theta$--morphism to (\ref{tree132}). We get
\beq
\ckappa{\theta}{}{\<\text{vac}|\theta^{(2)} ; {\bf v}\rangle}=
\<\text{vac}| \ckappa{\theta}{}{|\theta^{(2)} ; {\bf v}\rangle}=\<\text{vac}|{\bf v} \>+
g^2 \Gamma_{\bf v}\, \<\text{vac}| {\mathbb H}_{1,L_2} |0;{\bf v} \rangle\;.
\eeq
The first term in the right hand side is exactly what we are after. As for the last one we can use (\ref{PsiId}) with $j=L=L_2$ to get
\beq
\<\text{vac}|{\bf v} \>=\ckappa{\theta}{}{\<\text{vac}|\theta^{(2)} ; {\bf v}\rangle}-
g^2 \Gamma_{\bf v}  (i\partial_{1}-i\partial_{L_2} + \Gamma_{\bf v})\, \<\text{vac}|
\theta^{(2)};{\bf v} \rangle \la{almost}
\eeq
Note that in the right hand side of (\ref{tree132}) there is no dependence on $\theta_{L_2}$ and thus we can drop the $\partial_{L_2}$ from (\ref{almost}). All we have to do now is compute the derivatives in the right hand side of (\ref{almost}).
For that we use
\beqa
\ckappa{\theta}{}{{ e}^{L_{12}}_{\bar \alpha} } =(1-g^2 \Gamma_{\bar \alpha}^2) {\frak e}^{L_{12}}_{\bar \alpha}+ \mathcal{O}(g^4)\, , \qquad  g^2 i \partial_1 \({ e}^{L_{12}}_{\bar \alpha}  \) = -g^2 \Gamma_{\bar \alpha}{\frak e}^{L_{12}}_{\bar \alpha} + \mathcal{O}(g^4)
\eeqa
where we use the short-hand notation
\beq
{\frak e}^{L_{12}}_{\bar \alpha} \equiv \prod_{\bar a \in \bar\alpha} \(\frac{x(\bar a+i/2)}{x(\bar a-i/2)}\)^{L_{12}}\;.
\eeq
As usual, the $\Theta$--morphism naturally gives rise to the Zhukowsky variables. The obtained result
$\<\text{vac}|{\bf v} \>=(1- {g^2} \Gamma_{\bf v}^2)
\sum\limits_{\alpha \cup \bar\alpha=\bf v}
(-1)^{|\bar\alpha|}
f^{\alpha,\bar\alpha}{\frak e}^{L_{12}}_{\bar\alpha}\(1+{g^2} \Gamma_{\alpha}\Gamma_{\bar\alpha}\) $ can be simplified further to
\beqa
\<\text{vac}|{\bf v} \>
&=&(1- {g^2} \Gamma_{\bf v}^2)\sum_{\alpha \cup \bar\alpha=\bf v}
(-1)^{|\bar\alpha|}
{\frak f}^{\alpha,\bar\alpha}{\frak e}^{L_{12}}_{\bar\alpha} \,, \la{Aalg}
\eeqa
if we introduce a modified function
\beq
{\frak f}(u,v)\equiv \(1+\frac{i}{u-v}\)\(1+\frac{g^2}{(u^2+\tfrac14)(v^2+\tfrac14)}\)
\eeq
to absorb the partial energies $\Gamma_{\alpha}$ and $\Gamma_{\bar \alpha}$. The result we just \textit{derived} coincides precisely with the \textit{conjecture} in section \ref{Asec} based on the coordinate Bethe ansatz approach!\footnote{Of course, to compare both one needs to take (\ref{conv1}) into account.}

Using (\ref{Aalg}), (\ref{normPart1}), (\ref{BAEimpure2})  we can write the final result (\ref{C123oxo}) in the
following succinct form
\beq
 C_{123}^{\circ\bullet\circ} =
\frac{\displaystyle\sqrt{L_1L_2L_3 }}{\displaystyle\sqrt{\binom{L_1}{N_1}\binom{L_2}{N_3} }}
\frac{\displaystyle(1-{g^2} \Gamma_{\bf v}^2/2)\sum_{\alpha\cup\bar\alpha=\bf v}
(-1)^{|\bar\alpha|}
{\frak f}^{\alpha,\bar\alpha}{\frak e}^{L_{12}}_{\bar\alpha}}{\displaystyle\sqrt{ \prod_{i\neq j} f(u_i,u_j) \det_{j,k} \partial_{v_j} \phi_k^{\bf v}}}
\;,
\eeq
which agrees perfectly with the previously conjectured expression in the coordinate Bethe ansatz approach of section \ref{coorSec}. The case $ C_{123}^{\circ\bullet\bullet}$ is not harder to get, see \ref{decaysec}.

\subsection{Two non-BPS Three Point Function $C_{123}^{\bullet\bullet\circ}$  and also $C_{123}^{\bullet\bullet\bullet}$ }\la{sec73}
In this section we apply the  technology developed above
to the most complicated case of the structure constant when
all three operators are non-BPS. The operator ${\cal O}_3$ can be replaced by the corresponding BPS
operator without any loss of generality since its contribution factorizes and appears only in the \verb"simple" contribution in (\ref{tocompute}).
 Hence, we will first consider this to be the case and
 recover the general case at the very end.

Our starting point is again \eq{closed}. We pick two closed states at two loops
\beqa\la{closed2}
|{\bf 1}\rangle&\equiv&
(1-g^2 \Gamma_\bu {\mathbb H}_{L_1,1})
\ckappa{\theta^{(1)}}{ }{|\theta^{(1)};{\bf u}\rangle} \\
|{\bf 2}\rangle&\equiv&
(1-g^2 \Gamma_\bv {\mathbb H}_{L_2,1})
\ckappa{\theta^{(2)}}{ }{|\theta^{(2)};{\bf v}\rangle} \nn
\eeqa
Then we have to compute the most complicated contribution in (\ref{tocompute}) which is the last one, denoted as \verb"involved". We have
\beq
\verb"involved"=\<{\bf 1}|\hat{\mathcal{O}}_3|{\bf 2}\> -g^2 \<{\bf 1} | \(\mathbb{H}_{L_{12},L_{12}+1}+\mathbb{H}_{L_1,1}  \) \hat{\O}_3 \(\mathbb{H}_{L_{12},L_{12}+1}+\mathbb{H}_{L_2,1}  \) |{\bf 2}\>  \la{homework}
\eeq
where
\beq
\hat{\O}_3 = \sum_{i_1=\uparrow,\downarrow} \dots \sum_{i_{L_{12}}=\uparrow,\downarrow} \Big|i_1 \dots i_{L_{12}} \underbrace{\downarrow \dots \downarrow}_{L_{13}=N_3} \Big\> \Big\< i_1 \dots i_{L_{12}} \underbrace{\uparrow \dots \uparrow}_{L_{23}=L_3-N_3} \!\!\Big|  \,.
\eeq
It is also convenient to define the one loop states as $|{1}\> \equiv |0,{\bf u}\> $ and similar for $|{ 2}\>$. We shall not compute the second term in (\ref{homework}) since we will see soon that it neatly cancels with a contribution from the first one. Hence we focus on simplifying that first term
\beqa\la{startCxxo}
\<{\bf 1}|\hat{\mathcal{O}}_3|{\bf 2}\>&=&
\ckappa{\theta^{(1)}}{}{\langle\theta^{(1)};{\bf u}|}
\ckappa{\theta^{(2)}}{}{\hat O_3|\theta^{(2)};{\bf v}\rangle}
-g^2\langle 1|
( \Gamma_{\bu} {\mathbb H}_{L_1,1}\hat O_3
+\Gamma_\bv \hat O_3 {\mathbb H}_{L_2,1})
|2 \rangle\;.
\eeqa
Again we have the $\Theta$--morphism acting on the bra and ket states separately.
The strategy is as before: we want to to bring it to the form of
a differential operator in $\theta$'s acting on the tree level expression with impurities
\beq
 \<\theta^{(1)} ;{\bf u} | \hat{\mathcal{O}}_3| \theta^{(2)};{\bf v} \> \la{theguy} \,.
\eeq
This time it may seem a bit ambiguous since there are two different sets of $\theta$'s involved.
However, as explained in figure \ref{figC123}, the scalar product (\ref{theguy}) only depends on $\theta^{(1)}$ so that it is rather natural
to pick this set of $\theta$'s to act on. Moreover, the argument in that figure actually implies something slightly stronger, namely that the bottom part of the figure does not depend on $\theta^{(23)}$ and hence it only depends on $\theta^{(12)} \subset \theta^{(1)}$. The bottom is nothing but $\hat{\O}_{3} |\theta^{(2)};{\bf v}\rangle$ so we conclude that in (\ref{startCxxo}) the first term is\footnote{Very explicitly:
\beqa
\nn\ckappa{\theta^{(1)}}{}{\hat O_3|\theta^{(2)};{\bf v}\rangle}
&=&
\hat O_3|{\bf v}\rangle+\sum_{i=1}^{L_1}\frac{g^2}{2}\(\d_{\theta^{(1)}_{i}}-\d_{\theta^{(1)}_{i+1}}\)^2 {\hat O_3|
\theta^{(2)};{\bf v}\rangle}
\\
&=&
\hat O_3|{\bf v}\rangle+\frac{g^2}{2}\[
\sum_{i=1}^{L_{12}-1}\(\d_{\theta_{i}^{(1)}}-\d_{\theta^{(1)}_{i+1}}\)^2
+ \d_{\theta^{(1)}_{L_{12}}}^2
+ \d_{\theta^{(1)}_{1}}^2
\]
\ckappa{\theta^{(1)}}{}{\hat O_3|\theta^{(2)};{\bf v}\rangle}
=
\nn\ckappa{\theta^{(2)}}{}{\hat O_3|\theta^{(2)};{\bf v}\rangle}\;.
\eeqa}
\beqa\la{ourgoal}
\ckappa{\theta^{(1)}}{}{\langle\theta^{(1)};{\bf u}|}
\ckappa{\theta^{(2)}}{}{\hat O_3|\theta^{(2)};{\bf v}\rangle} = \ckappa{\theta^{(1)}}{}{\langle\theta^{(1)};{\bf u}|}
\ckappa{\theta^{(1)}}{}{\hat O_3|\theta^{(2)};{\bf v}\rangle}
\eeqa
By acting with the $\Theta$--morphism on (\ref{theguy}) we obtain the  right hand side of this expression plus crossed terms that need to be massaged as usual. This is done in detail in appendix \ref{derivationC123}. At the end of the day we find
\beqa\nn
\<{\bf 1}|\hat{\mathcal{O}}_3|{\bf 2}\>&=&\ckappa{\theta^{(1)}}{}{\<\theta^{(1)} ;{\bf u} | \hat{\mathcal{O}}_3| \theta^{(2)};{\bf v} \>}+
g^2(\Gamma_\bv-\Gamma_\bu)
\[i{\cal D}^{(1)}_{L_{12}}-\Gamma_\bv+\Gamma_\bu\]
\langle \theta^{(1)} ,{\bf u}|\nn\hat O_3|\theta^{(2)},{\bf v}\rangle
\\
&-&
g^2 (\Gamma_\bu+\Gamma_\bv
\Gamma_\bu
+\Gamma_\bv)\langle{1}|
\hat O_3
|{2}\rangle \nn \\
&+&g^2\langle{1}|
{\mathbb H}_{L_{12},1}\hat O_3-
{\mathbb H}_{L_1,1}\hat O_3-\hat O_3
{\mathbb H}_{L_{12},L_{12}+1}-
\hat O_3
{\mathbb H}_{L_2,1}|{2}\rangle\la{cc132sec}\;.
\eeqa
We see that the last line cancels precisely with the second term in (\ref{homework})! Recall that these terms came from internal Feynman loop contributions. We see that they neatly cancel with particular contributions coming from the wave function corrections.

More remarkable simplifications arise when we combine \verb"involved" with the other contributions in (\ref{tocompute}) to obtain the physical quantity $|C^{\bullet\bullet\circ}_{123}|$. For $\O_3$ BPS we have
\beq
|C^{\bullet\bullet\circ}_{123}|= \sqrt{\frac{L_1L_2L_3}{\binom{L_3}{N_3}}}\times |\verb"involved"/ \sqrt{\<{\bf 1}| {\bf 1}\>\<{\bf 2}| {\bf 2}\>}|
\eeq
Then dividing \eq{cc132sec} by the norms of  $|{\bf 1}\>$ and $|{\bf 2}\>$ given by \eq{normPart1} we get
\beqa
\la{almostthere}
\verb"involved"/ \sqrt{\<{\bf 1}| {\bf 1}\>\<{\bf 2}| {\bf 2}\>}&=&
\frac{
\ckappa{\theta^{(1)}}{}{\<\theta^{(1)} ;{\bf u} | \hat{\mathcal{O}}_3| \theta^{(2)};{\bf v} \>}
}{\sqrt{
\ckappa{\theta^{(1)}}{}{\langle\theta^{(1)};{\bf u}|\theta^{(1)};{\bf u}\rangle}
\ckappa{\theta^{(2)}}{}{\langle\theta^{(2)};{\bf v}|\theta^{(2)};{\bf v}\rangle}
}
}\\
&+&\frac{g^2(\Gamma_\bv-\Gamma_\bu)}{\sqrt{\<1|1\>\<2|2\>}}
\[i{\cal D}^{(1)}_{L_{12}}-\tfrac{1}{2}\(\Gamma_\bv-\Gamma_\bu\)\]
\<\theta^{(1)} ;{\bf u} | \hat{\mathcal{O}}_3| \theta^{(2)};{\bf v} \> \nn
\;.
\eeqa
As we show in Appendix \ref{impart} the last line is purely imaginary and thus
can be absorbed into a physically irrelevant phase prefactor in the left-hand side!
We conclude that the properly normalized structure constant is given by
\beq
|C^{\bullet\bullet\circ}_{123}|=\left|
\frac{\sqrt{L_1 L_2 L_3}
\ckappa{\theta^{(1)}}{}{\<\theta^{(1)} ;{\bf u} | \hat{\mathcal{O}}_3| \theta^{(2)};{\bf v} \>}
}{\sqrt{\binom{L_3}{N_3}
\ckappa{\theta^{(1)}}{}{\langle\theta^{(1)};{\bf u}|\theta^{(1)};{\bf u}\rangle}
\ckappa{\theta^{(2)}}{}{\langle\theta^{(2)};{\bf v}|\theta^{(2)};{\bf v}\rangle}
}
}
\right|\;.
\eeq
In sum, all ugly factors miraculously canceled out leaving us with the nicest possible expression!

From the discussion in section \ref{setup} we can easily generalize to the three BPS case
which only differs by an additional factor corresponding to the third operator:
\beq
|C^{\bullet\bullet\bullet}_{123}|=\left|
\frac{\sqrt{L_1 L_2 L_3}
\ckappa{\theta^{(1)}}{}{\<\theta^{(1)} ;{\bf u} | \hat{\mathcal{O}}_3| \theta^{(2)};{\bf v} \>}
\mathcal{A}_{N_3}({\bf w})(1-g^2\Gamma_{\bf w}/2)
}{\sqrt{
\ckappa{\theta^{(1)}}{}{\langle\theta^{(1)};{\bf u}|\theta^{(1)};{\bf u}\rangle}
\ckappa{\theta^{(2)}}{}{\langle\theta^{(2)};{\bf v}|\theta^{(2)};{\bf v}\rangle}
\<{\bf 3}|{\bf 3}\>
}
}
\right|\;. \la{super}
\eeq

Finally, it is straightforward to compute all derivatives appearing in the several $\Theta$--morphisms since we know the impure results, see section \ref{impure}. The final result for the one loop structure constants is presented in Appendix \ref{fullAfter}. The expression (\ref{super}) -- or equivalently the more explicit expressions in the Appendix \ref{fullAfter} -- is the main practical outcome of the $\Theta$--morphism.

\section{Conclusions} \la{Conc}
Let us conclude with a small list of the main ideas/results together with a link to the corresponding equations.

We constructed the eigenvectors of the two loop dilatation operator (\ref{dilatation0}) both in Coordinate
and Algebraic Bethe ansatz formalisms.
Using the Coordinate Bethe ansatz this amounts to finding all contact terms, see (\ref{contactTerms}) and (\ref{ctA}). In the Algebraic Bethe Ansatz approach we used the $\Theta$--morphism (\ref{mor1}) to promote the one loop eigenvectors
with extra parameters (impurities)
 (\ref{stateAlg}) into exact two loop
eigenvectors (\ref{closed}). In (\ref{Norm}) we conjectured the form of the norm of the states at {\it any loop order}
and made various tests of the conjectures by considering several states up to four loops.
We also give a derivation using the algebraic approach at two loops  in (\ref{normPart1}).

Another important scalar product is the Double Vacuum Decay (DVD) amplitude (\ref{Adef}),
which we found up to two loops in (\ref{Afin}) and (\ref{almost}). Using DVD as a building block
we found the structure constants $C_{123}^{\circ\bullet\bullet}$ (for $\O_1$ BPS)
 in (\ref{C123oxx}).

Finally, the $\Theta$--morphism techniques were used to derive the most general
structure constants $C_{123}^{\bullet\bullet\bullet}$ (all are non-BPS) in
  (\ref{super}) and (\ref{finalEvaluated}).

{  These expressions are basically ratios of  simple determinants.
As such they can be easily evaluated in a computer, even for operators with a large number of magnons which are the relevant operators in the so called classical limit.
 This allows for nice comparisons with strong coupling which were discussed in the note \cite{short}. }

It seems relatively straightforward to further correct the $\Theta$--morphism and take into account in this way the all loop corrections to the dispersion relation \cite{GSV}. In particular, we have full control over finite $L$ states. On the other hand, it seems harder to incorporate the dressing factor. It would be very interesting to make contact with the boost operator approach of \cite{boost} where the dressing factor corrections are under control for large $L$ states. To gain further insight into the underlying structure it is also very important to consider more general scalar products with fermions and derivatives.
We hope that with some more data, a unifying structure will emerge and very general all loop
expressions will become accessible.

\subsection*{Acknowledgments}
We thank J. Caetano, J. Escobedo, O. Foda, I. Kostov, D. Serban, A. Sever, J. Toledo, S. Valatka, T. Wang for discussions and suggestions.
N.G. (P.V.) would like to thank Nordita and the Perimeter Institute (King's College London, IHP and Nordita) for warm hospitality.
Research at the Perimeter Institute is supported in part by the Government of Canada through NSERC and by the Province of Ontario through MRI.

\newpage
\appendix
\section{The four magnon contact term} \la{C4}
By directly diagonalizing the two loop dilatation operator (\ref{dilatation0}) for large chains with four magnon excitations, we can derive the four particle contact term, see section \ref{contactSec}. We find $\mathbb{C}_{\bullet\bullet\bullet\bullet}^{(2)}(p_1,p_2,p_3,p_4)=2a/b$ where ($y_a=e^{ip_a}$)
\begin{eqnarray*}
a&=&
3 y_2^3 y_3^3 y_4^3 y_1^3-y_2^2 y_3^3 y_4^3 y_1^3-y_2^3 y_3^2 y_4^3 y_1^3+y_2^2 y_3^2 y_4^3 y_1^3-y_2^3 y_3^3 y_4^2 y_1^3+y_2^2 y_3^3 y_4^2 y_1^3+y_2^3 y_3^2 y_4^2 y_1^3\\
&-&4 y_2^2 y_3^2 y_4^2 y_1^3+y_2 y_3^2 y_4^2 y_1^3+y_2^2 y_3 y_4^2 y_1^3-y_2 y_3 y_4^2 y_1^3+y_2^2 y_3^2 y_4 y_1^3-y_2 y_3^2 y_4 y_1^3-y_2^2 y_3 y_4 y_1^3\\
&+&y_2 y_3 y_4 y_1^3-y_2^3 y_3^3 y_4^3  y_1^2+y_2^2 y_3^3 y_4^3 y_1^2+y_2^3 y_3^2 y_4^3 y_1^2-4 y_2^2 y_3^2 y_4^3 y_1^2+y_2 y_3^2 y_4^3 y_1^2+y_2^2 y_3 y_4^3 y_1^2\\
&-&y_2 y_3 y_4^3 y_1^2+y_2^2 y_3^2 y_1^2-y_2 y_3^2  y_1^2+y_2^3 y_3^3 y_4^2 y_1^2-4 y_2^2 y_3^3 y_4^2 y_1^2+y_2 y_3^3 y_4^2 y_1^2+y_2^2 y_4^2 y_1^2-4 y_2^3 y_3^2 y_4^2 y_1^2\\
&+&17 y_2^2 y_3^2 y_4^2 y_1^2-7 y_2 y_3^2 y_4^2 y_1^2+y_3^2  y_4^2 y_1^2-y_2 y_4^2 y_1^2+y_2^3 y_3 y_4^2 y_1^2-7 y_2^2 y_3 y_4^2 y_1^2+6 y_2 y_3 y_4^2 y_1^2\\
&-&y_3 y_4^2 y_1^2-y_2^2 y_3 y_1^2+y_2 y_3 y_1^2+y_2^2 y_3^3 y_4 y_1^2-y_2 y_3^3 y_4 y_1^2-y_2^2 y_4 y_1^2+y_2^3 y_3^2 y_4 y_1^2-7 y_2^2 y_3^2 y_4 y_1^2\\
&+&6 y_2 y_3^2 y_4 y_1^2-y_3^2 y_4 y_1^2+y_2 y_4 y_1^2-y_2^3 y_3 y_4 y_1^2+6 y_2^2 y_3 y_4 y_1^2-7 y_2 y_3 y_4  y_1^2+y_3 y_4 y_1^2+y_2^2 y_3^2 y_4^3 y_1\\
&-&y_2 y_3^2 y_4^3 y_1-y_2^2 y_3 y_4^3 y_1+y_2 y_3 y_4^3 y_1-y_2^2 y_3^2 y_1+y_2 y_3^2 y_1+y_2^2 y_3^3 y_4^2 y_1-y_2 y_3^3 y_4^2 y_1-y_2^2   y_4^2 y_1\\
&+&y_2^3 y_3^2 y_4^2 y_1-7 y_2^2 y_3^2 y_4^2 y_1+6 y_2 y_3^2 y_4^2 y_1-y_3^2 y_4^2 y_1+y_2 y_4^2 y_1-y_2^3 y_3 y_4^2 y_1+6 y_2^2 y_3 y_4^2 y_1-7 y_2 y_3 y_4^2 y_1\\
&+&y_3 y_4^2   y_1+y_2 y_1+y_2^2 y_3 y_1-4 y_2 y_3 y_1+y_3 y_1-y_2^2 y_3^3 y_4 y_1+y_2 y_3^3 y_4 y_1+y_2^2 y_4 y_1-y_2^3 y_3^2 y_4 y_1\\&+&
6 y_2^2 y_3^2 y_4 y_1-7 y_2 y_3^2 y_4 y_1+y_3^2 y_4 y_1-4   y_2 y_4 y_1+y_2^3 y_3 y_4 y_1-7 y_2^2 y_3 y_4 y_1+17 y_2 y_3 y_4 y_1-4 y_3 y_4 y_1\\
&+&y_4 y_1-y_1+y_2^2 y_3^2 y_4^2-y_2 y_3^2 y_4^2-y_2^2 y_3 y_4^2+y_2 y_3 y_4^2-y_2+y_2   y_3-y_3-y_2^2 y_3^2 y_4+y_2 y_3^2 y_4+y_2 y_4\\
&+&y_2^2 y_3 y_4-4 y_2 y_3 y_4+y_3 y_4-y_4+3
\end{eqnarray*}
and
\begin{eqnarray*}
b&=&y_2^3 y_3^3 y_4^3 y_1^3+y_2^2 y_3^2 y_4^3 y_1^3+y_2^2 y_3^3 y_4^2 y_1^3+y_2^3 y_3^2 y_4^2 y_1^3-2 y_2^2 y_3^2 y_4^2 y_1^3+y_2 y_3^2 y_4^2 y_1^3+y_2^2 y_3 y_4^2 y_1^3\\
&+&y_2^2 y_3^2 y_4   y_1^3+y_2 y_3 y_4 y_1^3+y_2^2 y_3^3 y_4^3 y_1^2+y_2^3 y_3^2 y_4^3 y_1^2-2 y_2^2 y_3^2 y_4^3 y_1^2+y_2 y_3^2 y_4^3 y_1^2+y_2^2 y_3 y_4^3 y_1^2+y_2^2 y_3^2 y_1^2\\
&+&y_2^3 y_3^3 y_4^2   y_1^2-2 y_2^2 y_3^3 y_4^2 y_1^2+y_2 y_3^3 y_4^2 y_1^2+y_2^2 y_4^2 y_1^2-2 y_2^3 y_3^2 y_4^2 y_1^2+7 y_2^2 y_3^2 y_4^2 y_1^2-4 y_2 y_3^2 y_4^2 y_1^2\\
&+&y_3^2 y_4^2 y_1^2+y_2^3 y_3  y_4^2 y_1^2-4 y_2^2 y_3 y_4^2 y_1^2+2 y_2 y_3 y_4^2 y_1^2+y_2 y_3 y_1^2+y_2^2 y_3^3 y_4 y_1^2+y_2^3 y_3^2 y_4 y_1^2-4 y_2^2 y_3^2 y_4 y_1^2\\
&+&2 y_2 y_3^2 y_4 y_1^2+y_2 y_4 y_1^2+2   y_2^2 y_3 y_4 y_1^2-4 y_2 y_3 y_4 y_1^2+y_3 y_4 y_1^2+y_2^2 y_3^2 y_4^3 y_1+y_2 y_3 y_4^3 y_1+y_2 y_3^2 y_1\\
&+&y_2^2 y_3^3 y_4^2 y_1+y_2^3 y_3^2 y_4^2 y_1-4 y_2^2 y_3^2 y_4^2 y_1+2 y_2 y_3^2 y_4^2 y_1+y_2 y_4^2 y_1+2 y_2^2 y_3 y_4^2 y_1-4 y_2 y_3 y_4^2 y_1+y_3 y_4^2 y_1\\
&+&y_2 y_1+y_2^2 y_3 y_1-2 y_2 y_3 y_1+y_3 y_1+y_2 y_3^3 y_4 y_1+y_2^2 y_4 y_1+2 y_2^2 y_3^2   y_4 y_1-4 y_2 y_3^2 y_4 y_1+y_3^2 y_4 y_1\\
&-&2 y_2 y_4 y_1+y_2^3 y_3 y_4 y_1-4 y_2^2 y_3 y_4 y_1+7 y_2 y_3 y_4 y_1-2 y_3 y_4 y_1+y_4 y_1+y_2^2 y_3^2 y_4^2+y_2 y_3 y_4^2+y_2 y_3\\&+&y_2   y_3^2 y_4+y_2 y_4+y_2^2 y_3 y_4-2 y_2 y_3 y_4+y_3 y_4+1
\end{eqnarray*}

\section{Simplifications for $\O_1$ BPS} \la{422ap}
In section \ref{decaysec} we explained that if $\O_1$ is a BPS operator we can simplify considerably the most complicated contribution \verb"involved". Here we provide some more details. When $\O_1$ is BPS,  \verb"involved"  is given by
\beqa
\left\<\,\ssststile{\,\,\,\,L_1\,\,\,\,}{\,\,N_1\,\,}\,\right|1-g^2( \mathbb{H}_{L_{12},L_{12}+1}+ \mathbb{H}_{L_1,1})
\left|\dots\ssststile{\,\,\,\,N_3\,\,\,\,}{\,\,N_3\,\,}\,\right\>\left\<\dots\ssststile{\,\,\,\,L_3-N_3\,\,\,\,}{\,\,0\,\,}\,\right|
1-g^2( \mathbb{H}_{L_{12},L_{12}+1}+ \mathbb{H}_{L_2,1})|{\bf 2}\> \nn
\eeqa
where the dots stand for all possible intermediate states of length $L_{12}$ (over which we sum).
We can drop the first two Hamiltonian insertions since they kill the bra. Then we use
$$\left\<\,\ssststile{\,\,\,\,L_1\,\,\,\,}{\,\,N_1\,\,}\,\right|\,
\left|\dots\ssststile{\,\,\,\,N_3\,\,\,\,}{\,\,N_3\,\,}\,\right\>\left\<\dots\ssststile{\,\,\,\,L_3-N_3\,\,\,\,}{\,\,0\,\,}\,\right| =\left\<\,\ssststile{\,\,\,\,L_1-N_3\,\,\,\,}{\,\,N_1-N_3\,\,}\,\,\ssststile{\,\,\,\,L_3-N_3\,\,\,\,}{\,\,0\,\,}\,\right| $$
Finally we use the same argument as in section \ref{simplificationSec} to replace $\mathbb{H}_{L_{12},L_{12}+1}+ \mathbb{H}_{L_2,1}$ by the full Hamiltonian which acts on the ket $|{\bf 2}\>$ to give the energy of the operator $\mathcal{O}_2$. We conclude that for $|{\bf 1}\>=\left|\,\ssststile{\,\,\,\,L_1\,\,\,\,}{\,\,N_1\,\,}\,\right>$ the \verb"involved" contribution
can be again written in terms of the DVD amplitude (\ref{Adef}), see (\ref{hereitis}).

\section{Code to generate contact terms} \la{codeContact}
Once we have guessed the form of the one loop scalar product (\ref{Adef}) there is a nice trick to use it to derive all contact terms. Indeed
\beq
\mathcal{A}_{N}({\bf p}) = \left\<\,\ssststile{\,\,\,\,N\,\,\,\,}{\,\,N\,\,}\,\ssststile{\,\,\,\,L-N\,\,\,\,}{\,\,0\,\,}\Big| \,{\bf p}\right\> = \psi(1,2,\dots,N) \la{Adef2}
\eeq
where $\psi(n_1,\dots,n_N)$ is the wave function and $L$ is the length of the Bethe state. Since all magnons are close together, it is given by
\beq
 \psi(1,2,\dots,N) =\(1+g^2\mathbb{C}^{(2)}_{\bullet\dots\bullet}(p_1,\dots,p_N) \)  \sum_{P} A(P)  \exp\(\sum_{j=1}^M j p_{Pj}\)
\eeq
where \beq
A(\dots, a,b, \dots) = \frac{u_a-u_b+i}{u_a-u_b-i} A(\dots, b,a,\dots) \,,
\eeq
and $A(1,2,\dots,M)=1$. Of course, $u_a=u(p_a)$ with $u(p)$ defined in (\ref{Smatrix}). Hence, we have
\beqa
g^2\mathbb{C}^{(2)}_{\bullet\dots\bullet}(p_1,\dots,p_N)  = 1-   \frac{ \displaystyle\sum_{P} A(P)  \exp\(\sum_{j=1}^M j p_{Pj}\)   }{\mathcal{A}_{N}({\bf p})} \la{ctA}
\eeqa
where $\mathcal{A}_{N}({\bf p})$ is  given by (\ref{Afin}). In this way we obtained a closed expression for all two loop contact terms. It would be interesting to generalize this to higher loops, see next section and \cite{GSV}.

In \verb"Mathematica",\footnote{The code becomes infinitely easier to read if converted into Standard Form. For that, copy the code into a Mathematica cell, select the cell, right click, and select "Convert To" and then "Standard Form".} \\ \\
\verb"e[u_,L_:1]=((I+2 u)/(-I+2 u))^L (1+(32 I g^2 L u)/(1+4 u^2)^2+O[g]^4);" \\
\verb"f=(1+I/(#1-#2))(1+g^2/((#1^2+1/4)(#2^2+1/4)))&;" \\
\verb"num[M_]:=Sum[A@@P Product[e[u[P[[j]]],j],{j,1,M}],{P,Permutations@Range@M}]//."\\
\verb"{A[a___,b_,c_,d___]:>(u[b]-u[c]+I)/(u[b]-u[c]-I)A[a,c,b,d]/;b>c,A@@Range[M]->1}" \\
\verb"den[M_]:=Product[1/(1/e[u[i]]-1)Product[1/f[u[i],u[j]],{j,i+1,M}],{i,1,M}]*"\\
\verb"Sum[Product[-e[j,M] Product[f[k,j],{k,a[[2]]}],{j,a[[1]]}]," \\
\verb"{a,{Complement[u/@Range[M],#],#}&/@Subsets[u/@Range[M],{0,M}]}]; " \\
\verb"Ct[M_]:=Ct[M]=1-num[M]/den[M]//SeriesCoefficient[#,{g,0,2}]&" \\ \\
For example, to get $\mathbb{C}_{\bullet\bullet\bullet}^{(2)}(p_1,p_2,p_3)$ we run
\verb"Ct[3]" (and \verb"Simplify" the result).

\section{Details on $\mathcal{A}$}\la{ApA}
\subsection{Two Magnon Computation}\la{ApA1}
Here we derive (\ref{A2}) and its one loop quantum corrected analogue in detail. We have
\beq
\la{state2} |{\bf p}\> =\sum_{1\le n_1 < n_2 \le L'} \psi(n_1,n_2)  |\overbrace{\underbrace{\uparrow \dots \uparrow}_{n_1-1}\! \downarrow \uparrow \dots\uparrow}^{n_2-1}\! \downarrow \uparrow \dots \>_{L'}
\eeq
where the subscript $L$ indicates the length of the state. The wave function is given by
\beqa \la{wave2}
 &&\psi(n_1,n_2)= \(e^{ip_1n_1+ip_2n_2}+e^{ip_2n_1+ip_1n_2} S(p_2,p_1) \)\times\\
 &&\qquad\qquad\qquad\qquad\qquad\(1+ \delta_{n_2,n_1+1} \,g^2\, \mathbb{C}_{\bullet\bullet}^{(2)}(p_2,p_1)+ \delta_{n_2=L',n_1=1} \,g^2\, \mathbb{C}_{\bullet\bullet}^{(2)}(p_2,p_1) \)\,.
 \nn\eeqa
 where we explicitly made manifest the last contact term contribution for the sake of clarity. Now
 \beq
\mathcal{A}_{L}({\bf p}) = \left\<\,\ssststile{\,\,\,\,L\,\,\,\,}{\,\,N\,\,}\,\ssststile{\,\,\,\,L'-L\,\,\,\,}{\,\,0\,\,}\Big| \,{\bf p}\right\>=\sum_{1\le n_1 < n_2 \le L} \psi(n_1,n_2) \la{state2open}
\eeq
Note that the sum goes only until $L<L'$.
The  only difference compared to (\ref{state2}) is the shortened length $L<L'$. Hence the last contact term in (\ref{wave2}) is never triggered.
Computing the sum (\ref{state2open}) using (\ref{wave2}) leads to (\ref{A2}) with $f(p,k)$ replaced by $\mathfrak{f}(p,k)$ in (\ref{fcorr}). Repeating the same exercise for a few cases with more magnons we find perfect agreement with (\ref{Afin}).

\subsection{Fixing $\mathfrak{f}$ from Impossibilities}\la{ApA2}
We have
\beq
\mathcal{A}_{L}({\bf p}) = \left\<\,\ssststile{\,\,\,\,L\,\,\,\,}{\,\,N\,\,}\,\ssststile{\,\,\,\,L'-L\,\,\,\,}{\,\,0\,\,}\Big| \,{\bf p}\right\>
 \eeq
 where $|{\bf p}\>$ is a Bethe state with rapidities ${\bf p}=\{p_1,\dots,p_N\}$ and length $L'$. To leading order this scalar product is given by (\ref{Asum}).

The scalar product $\mathcal{A}_L(\{p_1,\dots,p_N\})$ does not depend on $L'$, directly. That is, it is simply a function of the Bethe roots ${\bf p}$ and of the partial length $L$.\footnote{Of course, the momenta ${\bf p}$ do depend on $L'$ since they are the solution to Bethe equations on a circle of length $L'$. So the scalar product does depend on $L'$ in this sense. } It yields the probability amplitude of finding the last $L'-L$ sites of the chain empty and the $N$ excitations homogeneously distributed along the first $L$ sites. For this probability amplitude to be non-zero we obviously need
 \beq
 L\ge N \,. \la{ineq}
 \eeq
Very interestingly, the leading order expression  (\ref{Asum}) -- which was originally defined for $L\ge N$ only --  vanishes for $L<N$, see appendix $C$ of \cite{paper3}. This is a very nice fact which is of course not necessary a priori. It would be equally consistent -- but not as nice -- to have (\ref{Asum}) for $L\ge N$ only and simply define $\mathcal{A}_{L}=0$ for $L<N$.

 To leading order we have (\ref{Asum}) and we note that
 \beq
S(p,k)=\frac{f(p,k)}{f(k,p)}+\mathcal{O}(g^2)\,. \la{Sff}
 \eeq
In this appendix we propose an alternative path towards obtaining the higher loop generalization of (\ref{Asum}). We make the following assumptions:
\begin{itemize}
\item The scalar product continues to take the form (\ref{Asum}) and we simply need to correct $f(p,k)$ into $\mathfrak{f}(p,k)=f(p,k)+\mathcal{O}(g^2)$. This holds provided the lengths involved are large enough so that wrapping interactions do not play a role. That is, for safety we assume that $L'\gg L$ in the usual asymptotic Bethe ansatz sense.
\item At any loop order we have (\ref{Sff}), that is
\beq
S(p,k)=\frac{\mathfrak{f}(p,k)}{\mathfrak{f}(k,p)} \la{Sff2}
\eeq
\item We have
\beq
\mathcal{A}_{N-1}(p_1,\dots,p_N)=\mathcal{O}(g^{2N}) \,. \la{weaker}
\eeq
This is a generalization of $\mathcal{A}_{L}(p_1,\dots,p_N)=0$ for $L<N$ which is a property of (\ref{Asum}) to leading order as we discussed above. As mentioned above, this equality is some kind of a bonus that we did not have the right to ask for. As such, we will only require for a milder version of it to hold at higher loops, namely (\ref{weaker}). We might also impose other similar conditions for other values of $L<N$ but for the purpose of the current discussion we will only use (\ref{weaker}).\footnote{This condition somehow  resembles an asymptotic condition, that is of absence of "wrapping" corrections. }
\end{itemize}
We will now show how these assumptions fix the next loop order completely. For that consider the $N=2$ case. Using (\ref{Sff2}) we get rid of $\mathfrak{f}(p_2,p_1)$ in favor of $\mathfrak{f}(p_1,p_2)$ in $\mathcal{A}_L(p_1,p_2)$. Next we set $L=1$ and impose $\mathcal{A}_1(p_1,p_2)=\mathcal{O}(g^{4})$ following (\ref{weaker}). This fixes $\mathfrak{f}(p_1,p_2)$ to this loop order, see (\ref{A2}). We find
 \beq
\mathfrak{f}(p_1,p_2) = \frac{1 +e^{i  (p_1+p_2)} }{ e^{i  p_2}+S(p_2,p_1) e^{i p_1}}  \la{imp}
\eeq
which leads precisely to (\ref{fcorr}) which we obtained before by a completely different method! At the same time, $\mathcal{A}_{1}(p_1,\dots,p_N) = \mathcal{O}(g^2)$ for $N>2$. This confirms that at higher loops we can not ask for as much as for tree level. This is in agreement with the weaker condition (\ref{weaker}).

At the next few loop orders the conditions mentioned above seems to be valid. This statement is based on experimenting with configurations with small lengths and small number of magnons. For example,  the condition $\mathcal{A}_{2}(p_1,p_2,p_3)=\mathcal{O}(g^{6})$ seems to be indeed satisfied \cite{GSV}. On the other hand, this condition by itself is not enough to fix $\mathfrak{f}(p,k)$ completely.  By playing with more impossible configurations, with other $L$'s and $N$'s, it might be possible to constraint this function much more, eventually to all loop orders. That remains to be seen.

\section{Details of the derivation of the Structure Constant}\la{derivationC123}
In this appendix we present the derivation of (\ref{cc132sec}) in detail. As usual, the starting point is the action of the $\Theta$--morphism on an impure scalar product
\beqa\la{Cpe}
\ckappa{\theta^{(1)}}{}{\<\theta^{(1)} ;{\bf u} | \hat{\mathcal{O}}_3| \theta^{(2)};{\bf v} \>}&=&
\ckappa{\theta^{(1)}}{}{\langle\theta^{(1)};{\bf u}|}
\ckappa{\theta^{(1)}}{}{ \hat{\O}_3 |\theta^{(2)};{\bf v}\rangle}
+\cckappa{\theta^{(1)}}{}{\langle\theta^{(1)};{\bf u}|}{ \hat{\O}_3 |\theta^{(2)};{\bf v}\rangle}
\eeqa
The first term in the right hand side is exactly what we are after, see (\ref{ourgoal}), and the left-hand side is what we have already since we know (\ref{omar}). Hence we are left with the task of simplifying the cross-term.

As before we can recast the derivatives into  operators using \eq{PsiId}.
We use a shorthand notation ${\cal D}_i^{{(a)}}\equiv \d_{\theta^{(a)}_{i}}-\d_{\theta^{(a)}_{i+1}}$.
We also omit $\theta's$ in bras and kets as always $\theta^{(1)}$ appears in the combination
$\langle{\theta^{(1)};\bf u}|$ and similar for $\theta^{(2)}$
\beqa\la{crosss}
\nn&&\frac{1}{g^2}\cckappa{\theta_{1}}{}{\langle{\bf u}|}{ \hat{\O}_3 |{\bf v}\rangle}\!\!\!
=
\sum_{i=1}^{L_{1}}{\cal D}^{(1)}_i\langle{\bf u}|
 \hat{\O}_3 {\cal D}^{(1)}_i|{\bf v}\rangle\\
&&=
\sum_{i=1}^{L_{12}-1}{\cal D}^{(1)}_i\langle{\bf u}|
 \hat{\O}_3 {\cal D}^{(1)}_i|{\bf v}\rangle+
{\cal D}^{(1)}_{L_{12}}\langle{\bf u}|
 \hat{\O}_3 {\cal D}^{(1)}_{L_{12}}|{\bf v}\rangle
+
{\cal D}^{(1)}_{L_{1}}\langle{\bf u}|
 \hat{\O}_3 {\cal D}^{(1)}_{L_{1}}|{\bf v}\rangle\\
&&=
\sum_{i=1}^{L_{12}-1}{\cal D}^{(1)}_i\langle{\bf u}|
 \hat{\O}_3 {\cal D}^{(1)}_i|{\bf v}\rangle+
{\cal D}^{(1)}_{L_{12}}\langle{\bf u}|
 \hat{\O}_3 {\color{magenta}{\cal D}^{(2)}_{L_{12}}}|{\bf v}\rangle
+
{\cal D}^{(1)}_{L_{1}}\langle{\bf u}|
\nn \hat{\O}_3 {\color{magenta}{\cal D}^{(2)}_{L_{2}}}|{\bf v}\rangle
\eeqa
where we use again that $ \hat{\O}_3 \d_{\theta^{(1)}_{i}}|\bv\>= \hat{\O}_3 \d_{\theta^{(2)}_{i}}|\bv\>=0$
for $i>L_{12}$.
The first term can be simplified by converting the derivatives into operators
\beqa
&&\sum_{i=1}^{L_{12}-1}{\cal D}_i\langle{\bf u}|
 \hat{\O}_3 {\cal D}_i|{\bf v}\rangle =
\sum_{i=1}^{L_{12}-1}\langle{\bf u}|
{\mathbb H}_{i,i+1} \hat{\O}_3 {\mathbb H}_{i,i+1}|{\bf v}\rangle =
\sum_{i=1}^{L_{12}-1}\langle{\bf u}|
{\mathbb H}_{i,i+1} \hat{\O}_3 +
 \hat{\O}_3 {\mathbb H}_{i,i+1}|{\bf v}\rangle\\
\nn&&=
\langle{\bf u}|
\sum_{i=1}^{L_{1}}
{\mathbb H}_{i,i+1} \hat{\O}_3
+
\sum_{i=1}^{L_{2}}
 \hat{\O}_3  {\mathbb H}_{i,i+1}
-
{\mathbb H}_{L_{12},L_{12}+1} \hat{\O}_3 -
{\mathbb H}_{L_1,1} \hat{\O}_3 - \hat{\O}_3
{\mathbb H}_{L_{12},L_{12}+1}-
 \hat{\O}_3
{\mathbb H}_{L_2,1}|{\bf v}\rangle\;.
\eeqa
We use that all the Hamiltonian densities acting on the state $3$ give zero.
Note that the first two terms are the complete one loop Hamiltonians
acting on their eigenstates. Thus the first two terms are simply $\<\bu|(\Gamma_\bu+\Gamma_\bv) \hat{\O}_3 |\bv\>$.
Let us simplify the remaining two terms in \eq{crosss}. Using \eq{PsiId} again  we can rewrite them as
\beqa
\langle{\bf u}|
{\mathbb H}_{L_{12},L_{12}+1}
 \hat{\O}_3 {\mathbb H}_{L_{12},L_{12}+1}|{\bf v}\rangle
&+&
\langle{\bf u}|
({\mathbb H}_{L_1,1}-\Gamma_\bu)
 \hat{\O}_3
({\mathbb H}_{L_2,1}-\Gamma_\bv)
|{\bf v}\rangle\\
&=&\nn
\langle{\bf u}|
\Gamma_\bv
\Gamma_\bu
 \hat{\O}_3
-
\Gamma_\bv
{\mathbb H}_{L_1}
 \hat{\O}_3
-
\Gamma_\bu
 \hat{\O}_3
{\mathbb H}_{L_2}
|{\bf v}\rangle\;,
\eeqa
where we used that the terms with two $\mathbb H$ in the l.h.s. vanish. The reason is
that both Hamiltonian densities act with one leg on $ \hat{\O}_3 $. The Hamiltonian density is only nonzero
when the sites on which it acts are different, however when contracted with $ \hat{\O}_3 $ one of
them receives $\downarrow$ whereas another one $\uparrow$ and thus  one of the operators always gives zero
(see figure \ref{HH}).
\begin{figure}[t]
\begin{center}
\includegraphics[trim=14cm 15cm 0cm 0cm, clip=true, scale=1]{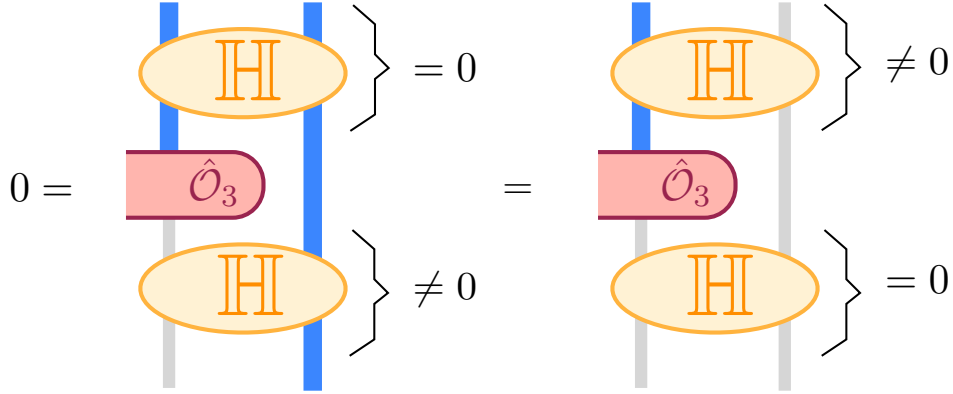}
\end{center}
\caption{As an operator $\mathbb{H}_{L_{12},L_{12}+1} \hat{\O}_3\mathbb{H}_{L_{12},L_{12}+1}  =\mathbb{H}_{L_{1},1} \hat{\O}_3\mathbb{H}_{L_{2},1}  =0$. The reason is that for the Hamiltonian density $\mathbb{H}=\mathbb{I}-\mathbb{P}$ to give a non-zero result it must act on two sites where the spins do not point in the same direction. If that is the case for the bottom (that is when acting on $\O_2$) then it will not be the case when acting on the top (that is when acting on $\O_1$). Similarly, if it acts on two different spins  in the top it will necessarily act on two equal spins in the bottom. These two cases are illustrated in the this figure. Hence we always get zero. } \la{HH}
\end{figure}
Thus the cross term finally gives
\beqa
&&\frac{1}{g^2}\cckappa{\theta_{1}}{}{\langle{\bf u}|}{ \hat{\O}_3 |{\bf v}\rangle}=\langle{\bf u}|
\Gamma_\bv
\Gamma_\bu
 \hat{\O}_3
-
\Gamma_\bv
{\mathbb H}_{L_1,1}
 \hat{\O}_3
-
\Gamma_\bu
 \hat{\O}_3
{\mathbb H}_{L_2,1}
|{\bf v}\rangle+\\
\nn&&
\langle{\bf u}|
\sum_{i=1}^{L_{1}}
{\mathbb H}_{i,i+1} \hat{\O}_3
+
\sum_{i=1}^{L_{2}}
 \hat{\O}_3  {\mathbb H}_{i,i+1}
-
{\mathbb H}_{L_{12},L_{12}+1} \hat{\O}_3 -
{\mathbb H}_{L_1,1} \hat{\O}_3 - \hat{\O}_3
{\mathbb H}_{L_{12},L_{12}+1}-
 \hat{\O}_3
{\mathbb H}_{L_2,1}|{\bf v}\rangle\;.
\eeqa
Combining all pieces together we get
\beqa
&&\<1|3|2\>-
\ckappa{\theta^{(1)}}{}{\<\theta^{(1)} ;{\bf u} | \hat{\mathcal{O}}_3| \theta^{(2)};{\bf v} \>}
=-
g^2\langle{\bf v}|
( \Gamma_\bu {\mathbb H}_{L_1,1} \hat{\O}_3
+\Gamma_\bv  \hat{\O}_3  {\mathbb H}_{L_2,1})
|{\bf v}\rangle\\
\nn&&-g^2\langle{\bf u}|
\Gamma_\bv
\Gamma_\bu
 \hat{\O}_3
-
\Gamma_\bv
{\mathbb H}_{L_1,1}
 \hat{\O}_3
-
\Gamma_\bu
 \hat{\O}_3
{\mathbb H}_{L_2,1}
|{\bf v}\rangle
\\
\nn&&-
g^2\langle{\bf u}|
(\Gamma_\bu+\Gamma_\bv) \hat{\O}_3
-
{\mathbb H}_{L_{12},L_{12}+1} \hat{\O}_3 -
{\mathbb H}_{L_1,1} \hat{\O}_3 - \hat{\O}_3
{\mathbb H}_{L_{12},L_{12}+1}-
 \hat{\O}_3
{\mathbb H}_{L_2,1}|{\bf v}\rangle
\eeqa
which simplifies to
\beqa\la{cc132}
\<1|3|2\>&=&\ckappa{\theta^{(1)}}{}{\<\theta^{(1)} ;{\bf u} | \hat{\mathcal{O}}_3| \theta^{(2)};{\bf v} \>}+
g^2(\Gamma_\bv-\Gamma_\bu)\langle{\bf u}|
{\mathbb H}_{L_1,1} \hat{\O}_3 -
 \hat{\O}_3 {\mathbb H}_{L_2,1}|{\bf v}\rangle
\\
&-&
g^2\langle{\bf u}|
(\Gamma_\bu+\Gamma_\bv
\Gamma_\bu
+\Gamma_\bv) \hat{\O}_3
-
{\mathbb H}_{L_{12},1} \hat{\O}_3 -
{\mathbb H}_{L_1,1} \hat{\O}_3 - \hat{\O}_3
{\mathbb H}_{L_{12},L_{12}+1}-
 \hat{\O}_3
{\mathbb H}_{L_2,1}|{\bf v}\rangle\nn\;.
\eeqa
The second term in the r.h.s. can be written in terms of a total derivative:
\beqa
\nn\langle{\bf u}|
{\mathbb H}_{L_1,1} \hat{\O}_3 -
 \hat{\O}_3 {\mathbb H}_{L_2,1}|{\bf v}\rangle &=&
\Big(i(\d_{\theta^{(1)}_{L_{12}}}-\d_{\theta^{(1)}_{L_{12}+1}})+\Gamma_\bu\langle{\bf u}|\Big)
 \hat{\O}_3 |{\bf v}\rangle\\
&+&
\langle{\bf u}|
 \hat{\O}_3 \Big(i(\d_{\theta^{(1)}_{L_{12}}}-0)-\Gamma_\bv|{\bf v}\Big)\\
&=& \[i{\cal D}^{(1)}_{L_{12}}-\Gamma_\bv+\Gamma_\bu\]\langle{\bf u}|
\nn \hat{\O}_3 |{\bf v}\rangle\;.
\eeqa
Finally we get
\beqa\la{cc132appendix}
\<1|3|2\>&=&\ckappa{\theta^{(1)}}{}{\<\theta^{(1)} ;{\bf u} | \hat{\mathcal{O}}_3| \theta^{(2)};{\bf v} \>}+
g^2(\Gamma_\bv-\Gamma_\bu)
\[i{\cal D}^{(1)}_{L_{12}}-\Gamma_\bv+\Gamma_\bu\]\langle{\bf u}|
\nn \hat{\O}_3 |{\bf v}\rangle
\\
&-&
g^2\langle{\bf u}|
(\Gamma_\bu+\Gamma_\bv
\Gamma_\bu
+\Gamma_\bv) \hat{\O}_3
-
{\mathbb H}_{L_{12},1} \hat{\O}_3 -
{\mathbb H}_{L_1,1} \hat{\O}_3 - \hat{\O}_3
{\mathbb H}_{L_{12},L_{12}+1}-
 \hat{\O}_3
{\mathbb H}_{L_2,1}|{\bf v}\rangle\nn\;.
\eeqa
Which is the expression used in section \ref{sec73}.

\subsection{On an imaginary term} \la{impart}
In this section we discuss the contribution of the second line in (\ref{almostthere}):
\beq
\frac{g^2(\Gamma_\bv-\Gamma_\bu)}{\sqrt{\<1|1\>\<2|2\>}}
\[i{\cal D}^{(1)}_{L_{12}}-\tfrac{1}{2}\(\Gamma_\bv-\Gamma_\bu\)\]
\<\theta^{(1)} ;{\bf u} | \hat{\mathcal{O}}_3| \theta^{(2)};{\bf v} \> \,, \qquad  {\cal D}^{(1)}_{L_{12}} = \frac{\partial}{\partial \theta^{(1)}_{L_{12}}}- \frac{\partial}{\partial \theta^{(1)}_{L_{12}+1}} \la{look}
\eeq
Now, from (\ref{omar}) we see that
\beq
\frac{\<\theta^{(1)} ;{\bf u} | \hat{\mathcal{O}}_3| \theta^{(2)};{\bf v} \>}{\<0 ;{\bf u} | \hat{\mathcal{O}}_3| 0;{\bf v} \>}= \frac{\prod_{m}^{N_3}\prod_{n}^{N_1}\(1-\frac{\hat\theta^{(1)}_{m}}{u_n+i/2}\)}{\prod_{m}^{N_3}\prod_{n}^{N_2}\(1-\frac{\hat\theta^{(1)}_{m}}{v_n+i/2}\)} \, \times \, \frac{1}{\prod_{m=1}^{N_2} \prod_{a=1}^{L_1} \frac{v_m-\theta_a -i/2}{v_m-i/2}} \, \times \verb"real"  \la{look2}
\eeq
The derivative $i {\cal D}^{(1)}_{L_{12}}$ can act on either of these three factors. After acting with the derivative we should set the impurities to zero since (\ref{look}) already has an overal $g^2$ factor. If $i {\cal D}^{(1)}_{L_{12}}$ acts on the real function it gives a purely imaginary contribution which does not contribute to the absolute value of the structure constants.\footnote{Recall that the structure constants are defined up to a phase. Multiplying an operator $\O_a$ by a phase does not affect its norm which is fixed by $\<\O_a \O^{\dagger}_a\>$ but it changes the three point function by that phase. When bootstrapping higher point functions these phases cancel out anyway since summing over what is flowing amounts to inserting $1=\sum_a |\O_a\>\<O_a|$ in the higher point function. Hence the phase of the intermediate operator drops out as it should.} If it acts on the second term it gives zero since the second term is a symmetric function of all the rapidities. Hence we are left with the actionof $i {\cal D}^{(1)}_{L_{12}}$ on the first factor of (\ref{look2}). That action leads to an imaginary part (which we drop as before) and to a real part. The real part  cancels precisely with the $\Gamma$ factors in (\ref{look})! Hence, the conclusion is simply that (\ref{look}) does not contribute to the absolute value of the structure constants.

\section{One Loop Structure Constants} \la{fullAfter}
After all dust settles\footnote{That is, after the derivatives in the $\Theta$--morphism are evaluated} we end up with the following explicit result for the one loop corrected structure constants:
\beq
C_{123}^{\bullet\bullet\bullet}({\bf u},{\bf v},{\bf w}) = \frac{\sqrt{L_1L_2L_3}\(1-g^2 \(\Gamma_{\bf w}+\Gamma_{\bf u \bf v}^2-\alpha_{\bf u\bf v}\)\)\mathcal{P}_{N_3}({\bf u},{\bf v}) \mathcal{A}_{N_3}({\bf w})\mathcal{S}_{N_3}({\bf u},{\bf v})}{\mathcal{B}({\bf w})  \mathcal{B}({\bf w}) \mathcal{B}({\bf v})} +\mathcal{O}(g^4) \,. \la{finalEvaluated}
\eeq
Let us summarize all ingredients in this expression to render this appendix self-contained. The three operators are parametrized by $3$ sets of Bethe roots: ${\bf u}=\{u_{j=1,\dots,N_1}\}$, ${\bf v}=\{v_{j=1,\dots,N_2}\}$ and ${\bf w}=\{w_{j=1,\dots,N_3}\}$. Then, starting from the simplest contributions, we have
\beq
\Gamma_{\bf w} \equiv \sum_{j=1}^{N_3} \frac{1}{w_j^2+\frac{1}{4}} \, , \qquad \Gamma_{\bf uv} \equiv \frac{1}{2}\sum_{j=1}^{N_1} \frac{1}{u_j^2+\frac{1}{4}}-\frac{1}{2}\sum_{j=1}^{N_2} \frac{1}{v_j^2+\frac{1}{4}} \, , \qquad \alpha_{\bf uv} \equiv \sum_{j=1}^{N_1} \frac{u_j}{u_j^2+\frac{1}{4}}- \sum_{j=1}^{N_2} \frac{v_j}{v_j^2+\frac{1}{4}} \,.  \nn
\eeq
Then
\beq
\mathcal{P}_{N_3}({\bf u},{\bf v}) \equiv \frac{\prod\limits_{j=1}^{N_2} \frac{Q(v_j)}{(g x(v_j-i/2))^{N_3}} \prod\limits_{j=1}^{N_1}  (g x(u_j-i/2))^{N_3}}{\prod\limits_{j<k}(u_j-u_k+i)\prod\limits_{j<k}(v_k-v_j+i)} \nn \qquad  \text{with} \qquad Q(z)\equiv \prod_{j=1}^{N_1} (z-u_j)
\eeq
and where the famous Zhuwkosky variables
\beq
x(u)+\frac{1}{x(u)} \equiv \frac{u}{g} \,, \qquad x(u) = \frac{u}{g}-\frac{g}{u}+\mathcal{O}(g^3) \,. \nn
\eeq
We now get to the interesting factors. We have
\beq
\mathcal{A}_{N_3}({\bf w}) \equiv \frac{ \displaystyle \sum_{\alpha \cup \bar \alpha = \bf w} (-1)^{\alpha}\! \! \prod_{a \in \alpha, \bar a \in \bar \alpha} \! \! \mathfrak{f}(a,\bar a) \,\prod_{\bar a \in \bar\alpha} \(\frac{x(\bar a+i/2)}{x(\bar a-i/2)}\)^{N_3}  }{\displaystyle\prod\limits_{j=1}^{N_3} \(\frac{x(w_j-i/2)}{x(w_j+i/2)}-1\) \prod\limits_{1\le i<j \le N} \mathfrak{f}(w_i,w_j)} \la{Acor}
\eeq
with
\beq
\mathfrak{f}(w,z)\equiv \(1+\frac{i}{w-z}\)  \(1+\frac{g^2}{(w^2+\frac{1}{4})(z^2+\frac{1}{4})} + \mathcal{O}(g^4)\) \,. \nn \eeq
Then we have the denominators
\beq
\mathcal{B}({\bf u}) \equiv \sqrt{ \sqrt{ \prod_{n<m} \frac{S(u_n,u_m)}{S(u_n^*,u_m^*)}} \det_{1\le i,j\le N_1} \frac{\partial}{\partial u_j}\( \frac{L}{i} \log\frac{x(u_k+i/2)}{x(u_k-i/2)}+\sum_{l\neq k} \frac{1}{i} \log \frac{u_k-u_l-i}{u_k-u_l+i}\) } \nn
\eeq
with similar expressions for $\mathcal{B}({\bf u})$ and $\mathcal{B}({\bf w})$. The S-matrix
\beq
S(u,v) \equiv \frac{v-u+i}{v-u-i} \,. \nn
\eeq
We are left with the last, most involved structure. We have
\beq
\mathcal{S}_{N_3}({\bf u},{\bf v}) \equiv \mathcal{D}+g^2\Big((N_3+1) \mathcal{D}^{[+2]}+(N_3-1) \mathcal{D}^{[+1,+1]}-2 \alpha_{\bf u \bf v} \mathcal{D}^{[+1]} \Big) \la{doesitfit}
\eeq
where
\beqa
\mathcal{D}&\equiv& \left|\begin{array}{ccccccc}
\frac{\partial \mathcal{T}(v_1)}{\partial u_1} & \dots &  \frac{\partial \mathcal{T}(v_{N_2})}{\partial u_1} & q_2(u_1) & \dots & q_{N_3}(u_1) & q_{N_3+1}(u_1)\\
\vdots & & \vdots & \vdots & &\vdots & \vdots \\
\frac{\partial \mathcal{T}(v_1)}{\partial u_{N_1}} & \dots &  \frac{\partial \mathcal{T}(v_{N_2})}{\partial u_{N_1}} & q_2(u_{N_1}) & \dots &q_{N_3}(u_{N_1})& q_{N_3+1}(u_{N_1})
\end{array}
\right| \nn\\
\mathcal{D}^{\color{red}[+2]}&\equiv& \left|\begin{array}{ccccccc}
\frac{\partial \mathcal{T}(v_1)}{\partial u_1} & \dots &  \frac{\partial \mathcal{T}(v_{N_2})}{\partial u_1} & q_2(u_1) & \dots &q_{N_3}(u_1) & q_{N_3+1{\color{red}+2}}(u_1)\\
\vdots & & \vdots & \vdots & & \vdots \\
\frac{\partial \mathcal{T}(v_1)}{\partial u_{N_1}} & \dots &  \frac{\partial \mathcal{T}(v_{N_2})}{\partial u_{N_1}} & q_2(u_{N_1}) & \dots &q_{N_3}(u_{N_1})& q_{N_3+1{\color{red}+2}}(u_{N_1})
\end{array}
\right| \nn\\
\mathcal{D}^{\color{red}[+1,+1]}&\equiv& \left|\begin{array}{ccccccc}
\frac{\partial \mathcal{T}(v_1)}{\partial u_1} & \dots &  \frac{\partial \mathcal{T}(v_{N_2})}{\partial u_1} & q_2(u_1) & \dots &q_{N_3{\color{red}+1}}(u_1) & q_{N_3+1{\color{red}+1}}(u_1)\\
\vdots & & \vdots & \vdots & & \vdots \\
\frac{\partial \mathcal{T}(v_1)}{\partial u_{N_1}} & \dots &  \frac{\partial \mathcal{T}(v_{N_2})}{\partial u_{N_1}} & q_2(u_{N_1}) & \dots &q_{N_3{\color{red}+1}}(u_{N_1})& q_{N_3+1{\color{red}+1}}(u_{N_1})
\end{array}
\right| \nn\\
\mathcal{D}^{\color{red}[+1]}&\equiv& \left|\begin{array}{ccccccc}
\frac{\partial \mathcal{T}(v_1)}{\partial u_1} & \dots &  \frac{\partial \mathcal{T}(v_{N_2})}{\partial u_1} & q_2(u_1) & \dots &q_{N_3}(u_1) & q_{N_3+1{\color{red}+1}}(u_1)\\
\vdots & & \vdots & \vdots & & \vdots \\
\frac{\partial \mathcal{T}(v_1)}{\partial u_{N_1}} & \dots &  \frac{\partial \mathcal{T}(v_{N_2})}{\partial u_{N_1}} & q_2(u_{N_1}) & \dots &q_{N_3}(u_{N_1})& q_{N_3+1{\color{red}+1}}(u_{N_1})
\end{array}
\right| \nn
\eeqa
Finally,
\beq
q_n(u)\equiv \frac{i}{(u+i/2)^{n-1}}-\frac{i}{(u+i/2)^{n-1}}  \,\,\,\,\,\,\, \text{and} \,\,\,\,\,\,\, \mathcal{T}(u) \equiv \frac{Q(u-i)}{Q(u)} + \(\frac{x(u-i/2)}{x(u+i/2)}\)^L \frac{Q(u+i)}{Q(u)}   \,. \nn
\eeq
We end with some comments.
\begin{itemize}
\item For symmetric configurations of Bethe roots $\alpha_{\bf u\bf v} = 0$ and the above expressions simplify slightly.
\item If $\O_3$ is BPS the structure constant $C_{123}^{\bullet\bullet\bullet}$ reduces to $C_{123}^{\bullet \bullet \circ}$. This case \textit{is not} considerably simpler than the general case. To obtain this case we replace $\Gamma_{\bf w} \to 0$, $\mathcal{B}({\bf w})\to \sqrt{\binom{L_3}{N_3}}$ and $\mathcal{A}_{N_3}({\bf w})\to 1$ in (\ref{finalEvaluated}).
\item If $\O_1$ is BPS the structure constant $C_{123}^{\bullet\bullet\bullet}$ reduces to $C_{123}^{\circ \bullet \bullet}$. This case \textit{is} considerably simpler than the general case. The quantity $\mathcal{S}_{N_3}$ does not appear in this case, only
$\mathcal{A}$ type quantities arise. This case is discussed in the main text, see section \ref{decaysec}.
\item The contribution (\ref{Acor}) can  be written as a determinant \cite{omar2}.
\end{itemize}

\end{document}